\makeatletter
\@ifundefined{@parse@version@dash}{%
\def\@parse@version#1{\@parse@version@0#1}
\def\@parse@version@#1/#2/#3#4#5\@nil{%
\@parse@version@dash#1-#2-#3#4\@nil}
\def\@parse@version@dash#1-#2-#3#4#5\@nil{%
  \if\relax#2\relax\else#1\fi#2#3#4 }
}{}
\makeatother

\documentclass[aps, 
 longbibliography,
 prfluids, 
 reprint,
 superscriptaddress,
 onecolumn,
 amsmath,amssymb
]{revtex4-2}

\usepackage{graphicx, import}
\usepackage{dcolumn}
\usepackage{bm}
\usepackage{siunitx}
\DeclareSIUnit\Molar{\textsc{m}}
\usepackage{tikz,pgfplots}

\newlength\figureheight
\newlength\figurewidth

\pgfplotsset{compat=newest}
\usepackage{tikzscale}
\usepgfplotslibrary{units}
\usepackage{textgreek}
\usepackage{amsmath}
\usepackage{cases}
\usepackage{epstopdf}

\DeclareUnicodeCharacter{2212}{-}

\usepackage{stmaryrd} 

\makeatletter
\renewcommand*{\@fnsymbol}[1]{\ensuremath{\ifcase#1\or \S \or *\or \ddagger\or
    \mathsection\or \mathparagraph\or \|\or **\or \dagger\dagger
    \or \ddagger\ddagger \else\@ctrerr\fi}} 
\makeatother

\begin{document}


\title{Deformation modes of an oil-water interface under a local electric field: \\From Taylor cones to surface dimples}

\author{Sebastian Dehe}
\affiliation{%
 Department of Mechanical Engineering, Technische Universit{\"a}t Darmstadt, 64287 Darmstadt, Germany
}%


\author{Steffen Hardt}
\email[corresponding author: ]{S.H. (hardt@nmf.tu-darmstadt.de)}
\affiliation{%
 Department of Mechanical Engineering, Technische Universit{\"a}t Darmstadt, 64287 Darmstadt, Germany
}%


\date{\today}

\begin{abstract}
Fluidic interfaces disintegrate under sufficiently strong electric fields, leading to electrohydrodynamic (EHD) tip streaming. Taylor cones, which emit charged droplets from the tip of a conical cusp, are among the most prominent and well-studied examples of EHD instabilities. 
In liquid-liquid systems, more complex interface deformation modes than simple Taylor cones can be observed, with the interface being pushed away from the electrode, and additional cone structures emerging from the rim of the dimple. 
In this article, we investigate the mechanisms behind these deformation modes experimentally and numerically, and demonstrate that the presence of droplets triggers the dimple at the interface.
In order to characterize the underlying processes, we replace the pin electrode by a hollow metallic needle with a prescribed electrolyte volume flow. The submerged electrospray introduces droplets of an aqueous KCl solution with varying ion concentrations into silicone oils with varying viscosities. By measuring the corresponding electric current and by optical investigation of the interface deformation, we study the system response to variations of the ionic concentration, viscosity, applied voltage as well as flow rate. 
The voltage between needle and liquid-liquid interface has a strong influence on the deformation, whereas the electrolyte flow rate only has a small influence. In addition, we observe that both the deflection as well as the current reach limiting values with increasing ion concentration and viscosity, which we explain based on a scaling relationship for the size of the droplets. 
In addition to the experiments, we use a finite element solver and compute the charge transport due to the droplets in the oil phase as an advection-diffusion process. Further, we compute the electric potential distribution, flow field and interface deformation. After calibration of our model with particle tracking velocimetry data of the flow inside the oil phase, we reproduce the experimentally observed dimple at the liquid-liquid interface. By including a relationship between the electric field at the pin electrode and the emitted electrospray current, we are able to identify the space charge as the cause of the experimentally observed limiting behavior of the interface deflection and the electric current. In summary, this work highlights the importance of charged droplets for the complex dynamic modes observed when a liquid-liquid interface is exposed to a local electric field. 
\end{abstract}

                              
\maketitle

\section{Introduction}

The disintegration of fluidic interfaces under electric fields has been of interest to scientists for more than a century \cite{Zeleny1917, Wilson1925, Taylor1964}. As a reaction to an applied electric field, an interface between a dielectric and a conducting fluid forms a conical shape, today referred to as Taylor cone, and exhibits an instability, resulting in the ejection of a fine jet or droplets once a critical field strength is exceeded.
The breakup is governed by a balance of electric, capillary and hydrodynamic forces, thus leading to the terminology of electrohydrodynamic (EHD) tip streaming \cite{Saville1997, Collins2008}. 
EHD tip streaming is utilized in a wide range of applications, such as liquid atomization \cite{Grace1994}, mass spectrometry \cite{Fenn1989}, electrospinning \cite{Teo2006} and printing \cite{Park2007}. 

As a result of the widespread applications of EHD tip streaming, an extensive range of operating regimes has been investigated \cite{Jaworek1998, DelaMora2007}.
For example, a large body of work exists on the steady cone-jet mode, as it allows to produce monodisperse micrometer sized droplets, while simultaneously yielding a reproducible emitted current and mass flow rate \cite{Rosell-Llompart1994, Hartman1999,Ponce-Torres2018}.
However, other operational modes involving whipping \cite{Riboux2011}, electrodripping \cite{Marginean2006a}, and pulsating cones \cite{Marginean2006, Bober2011}  as well as variations including coaxial jets \cite{Loscertales2002, Lopez-Herrera2003}, flow-focusing \cite{Ganan-Calvo2006,Kim2007} and multiple Taylor cones \cite{Bocanegra2005} have been of interest as well.
In that context, the influence of the material properties of the liquid being sprayed, such as the conductivity \cite{Tang1991,DelaMora1994,Tang1996} and viscosity \cite{Ku2002,Higuera2010}, has been studied, as well as the influence of the outer medium, including vacuum \cite{Gamero-Castano2008} as well as various dielectric liquids \cite{Gundabala2010,Marin2012}.
Commonly, for technological applications a fluid flow is supplied by external pumping. However, the tip streaming from droplets \cite{Collins2013, Pillai2016} as well as from liquid films \cite{Collins2008} can yield insights into the physical mechanisms, for example to derive or validate scaling models for the droplet size and charge.

As EHD tip streaming is present in a range of sophisticated technological applications, a thorough understanding of the process is crucial. A prototypic experimental setup to study EHD tip streaming is a pin electrode above a fluidic interface, due to its well-controllable parameters and its simplicity. For example, Collins and coworkers \cite{Collins2008} investigated the tip-streaming from a planar film and demonstrated the existence of a fundamental scaling law for the charge of a droplet pinched off from a Taylor cone. Pillai and coworkers extended the analysis to liquid-liquid systems by means of numerical simulations of droplets and observed related scaling laws \cite{Pillai2016}. However, experiments where an electric field is applied to a liquid-liquid interfaces can lead to a different, much less reported phenomenon. Instead of being attracted towards the electrode and forming a Taylor cone, the interface is repelled by the electrode and forms a dimple, with cones emerging from its rim. More interestingly, both configurations, the classical Taylor cone and the dimple, can be observed in the same system at similar values of the applied voltage, when the experiment is repeated a number of times. As we will discuss, the existence of droplets of the conductive phase in the dielectric liquid plays a crucial role in determining which configuration will be observed. By utilizing analogies to submerged electrosprays, we provide an explanation for the occurrence of dimples in the liquid-liquid interface. The existence of droplets leads to memory effects in the system response, and their diameter can be well below the optical resolution of the experiment. As a result, the observed dynamics and configurations could potentially be falsely attributed to other effects, underlining the importance to thoroughly understand these effects.

The paper is organized as follows: In section \ref{sec:exp_setup}, we discuss the experimental configurations utilized to study EHD tip streaming. In section \ref{sec:pin_interface_experiments}, we highlight the different breakdown mechanisms of the liquid-liquid interface under the electric field due to a pin-electrode. Section \ref{sec:needle_interface_configuraion} reports about a modified setup such that droplets can be injected from a metallic needle in form of a submerged electrospray. By characterizing the influence of the system parameters on the dimple at the liquid-liquid interface, we identify the role the droplets play for the dynamics of the system. In section \ref{sec:numerical_modeling}, we corroborate our hypothesis by numerical computations of the electric potential, fluid flow and charge transport, reproducing the experimental results for the dimple at the interface. Finally, in section \ref{sec:conclusion}, we discuss and summarize our findings.

\section{Experimental Setup}
\label{sec:exp_setup}
\begin{figure}[!htb]
\centering
\includegraphics[]{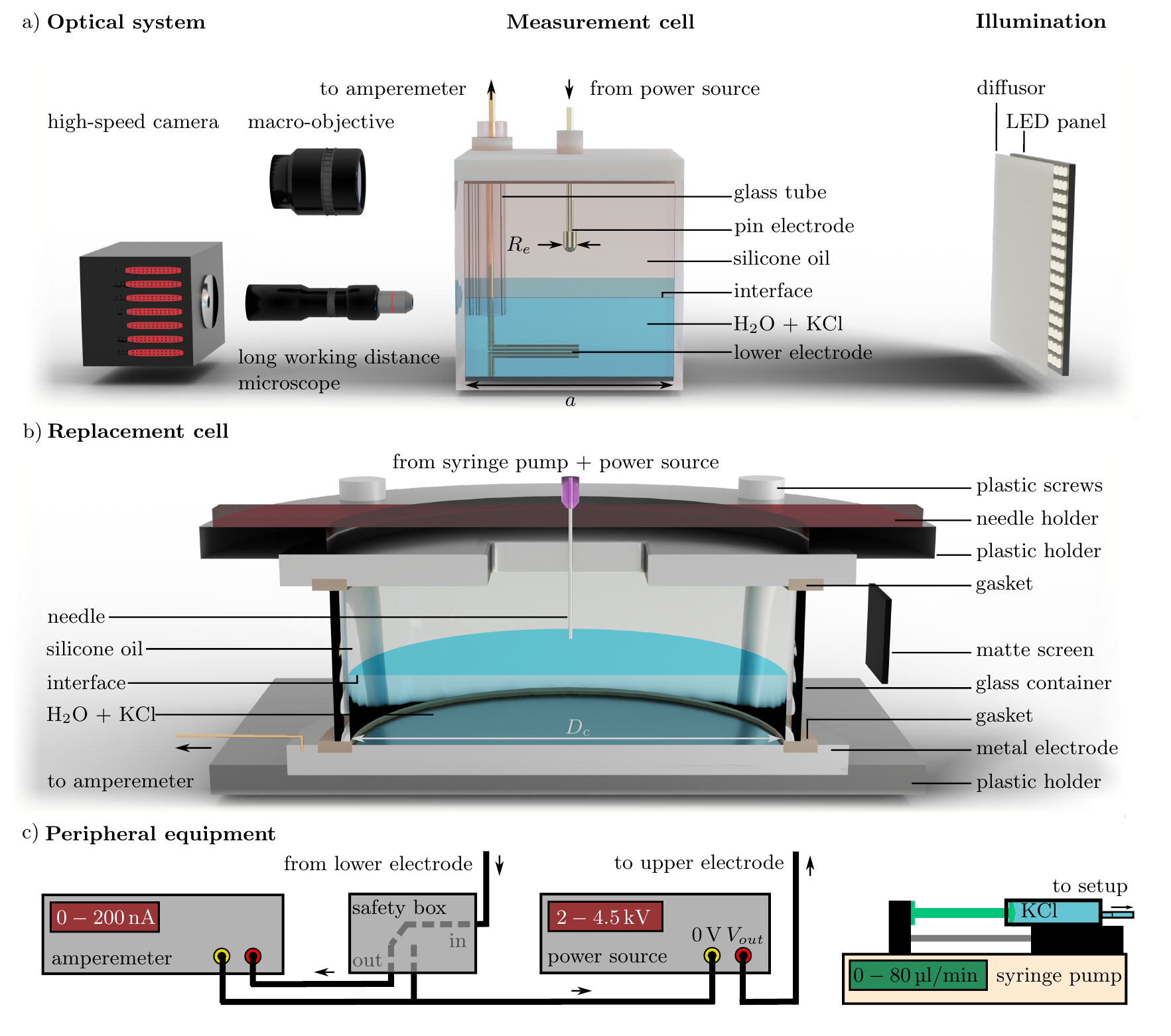}
\caption{\label{fig:Exp_setup} Schematic of the experimental setups used to study liquid-liquid interfaces under electric fields and the effect of droplets onto the system. 
a) The experimental setup consists of a pin electrode placed within a glass container of square base with $a=\SI{36}{\milli \meter}$, above a liquid-liquid interface. The upper phase is a dielectric liquid (e.g. silicone oil) and the lower one an aqueous KCl solution that is contacted with a second electrode. The interface is observed utilizing a high-speed camera with either a DSLR-camera macro-objective, or a long working distance microscope, and illuminated by a LED light panel from the back. 
b) Replacement cell used to introduce droplets into the system in a controlled manner. It consists of a cylindrical container with a diameter $D_c=\SI{12.4}{\centi\metre}$. Droplets are ejected from the hollow needle at the top with fixed volume flow of an aqueous KCl solution. The cell is sealed using a clamping plastic holder.
c) For both setups, the voltage is controlled by a high-voltage power source. The electric current is recorded by a pico-ammeter that is protected from overvoltage by a safety box. The flow rate is controlled by utilizing a syringe pump.
}
\end{figure}

In order to study the dynamics of liquid-liquid interfaces under electric fields, we designed two distinct experimental setups (see Fig. \ref{fig:Exp_setup}(a-b)). First, we utilize a glass container (\textit{Kr\"{u}ss}) of rectangular base with an inner side length of $a=\SI{36}{\milli \meter}$, as illustrated in Fig. \ref{fig:Exp_setup}(a). We position the pin electrode using a 3D-printed lid with an adjustable clamping mechanism to vary the distance between the electrode and the liquid-liquid interface. Also, the electrode geometry can be varied by replacing the electrode. To electrically contact the lower conducting phase, we pass a second electrode through a glass tube from the upper side of the lid to the bottom of the cell. With the upper liquid being a dielectric medium with negligible conductivity, this configuration ensures that no current is transported through the upper phase. We fill the container through an additional port at the upper lid \textit{via} a pipette and 
adjust the layer thicknesses $h_w$, $h_o$ of the lower and the upper layer, respectively, as desired. 

A high-speed camera (\textit{FASTCAM Mini AX200, Photron}) records the liquid-liquid interface from the side with varying framerates, utilizing either a DSLR objective (\textit{AF-S Micro NIKKOR 105mm 1:2,8G VR, Nikon}) or a long-working distance microscope ($6\times$ magnification, \textit{Navitar}). We place the camera on a mount (not shown) with adjustable height and incident angle with respect to the glass container. Tilting the camera allows to observe phenomena occurring close to the interface, which otherwise would be blocked by the mensicus at the side wall of the container. We control the camera \textit{via} computer using an ethernet connection and the proprietary software \textsc{Photron Fastcam Viewer 4}. An LED panel with a diffusor sheet and adjustable intensity illuminates the setup. We apply the electric potential difference $\phi_n$ between the electrodes with a high-voltage power source (\textit{Heinzinger PNC-series $\SI{6}{\kilo V}$} or \textit{Labsmith HVS448 – 6000D}, depending on the experiment). A picoammeter (\textit{Keithley 6485}) monitors the current flowing from the lower electrode to the power source, while a self-designed safety box prevents operational conditions outside the device specifications (see Fig. \ref{fig:Exp_setup}(c) for overview over peripheral devices \cite{eiffert2015}). We confirmed that the measurements were not influenced by the safety box by repeating current measurements without the safety box with an otherwise unchanged setup, leading to no significant difference.

Controlling the number of droplets of the conductive phase dispersed in the oil phase is critical for the experiments designed to explain the formation of a dimple in the liquid-liquid interface. Therefore, we designed a second experimental cell (Fig. \ref{fig:Exp_setup}(b)), while maintaining the optical and peripheral devices. The cell consists of a circular glass cylinder with an inner diameter of $D_c = \SI{12.4}{\centi \meter}$, which is fixed at the top and bottom by circular plates with embedded gaskets. The cell size is significantly increased in order to suppress wall effects. The lower plate serves as the bottom electrode, and the upper plate has a square opening for filling and electrode placement. A plastic holder clamps both plates with plastic screws, in order to prevent electrical contact to the environment. The upper electrode is held in place by a second plastic holder with an adjustable clamping mechanism, but in contrast to the first setup, a hollow metallic needle (inner diameter $\SI{0.58}{\milli \meter}$, outer diameter $\SI{0.91}{\milli \meter}$, length $\SI{3.81}{\centi \meter}$, \textit{Vieweg}) is used as electrode. By contacting the needle with a wire we connect it to the power source. We control the volume flow $Q$ of an electrolyte fed to the needle using a syringe pump (\textit{KD-Scientific KDS-210-CE}) connected by plastic tubing. Thereby, we are able to introduce droplets into the upper phase and study their effects on the system behavior. Additionally, a matte black screen on the back of the cell prevents light from entering the region of interest directly. As a result, the interface reflects light from the LED panel that has a significantly larger size than the screen, and thus the interface is well visible as a bright region in front of a dark background (see Fig. \ref{fig:Dip_observation} for more details). 

\section{Basic configurations and dynamics of the liquid-liquid interface}
\label{sec:pin_interface_experiments}

When a dielectric-electrolyte interface is exposed to an electric field using a pin electrode placed a few $\si{\milli \metre}$ above the interface, it disintegrates when a critical voltage is exceeded.
The Maxwell stress due to electrostatic forces enters the Navier-Stokes equation for an incompressible fluid as a force term of the form \cite{Saville1997}
\begin{equation}
\label{eq:Maxwell_stress}
\vec{f}_M = \nabla\cdot\sigma^M 
=\nabla \cdot \left[ \epsilon_0 \epsilon_\text{rel} \vec{E}\otimes\vec{E}-\frac{1}{2}\epsilon_0 \epsilon_\text{rel} \left(\vec{E}\cdot \vec{E} \right) \mathbb{I}\right],
\end{equation}
where $\sigma^M$ denotes the Maxwell stress tensor, $\vec{E}$ the electric field, $\epsilon_0$ the vacuum permittivity, $\epsilon_\text{rel}$ the relative permittivity of the liquid, and $ \mathbb{I}$ the identify tensor.
The first term represents the influence of polarization, while the second term represents the Coulomb force on the free charge. At a fluid interface, the free electric charge per unit area $q$ leads to a jump of the electric field as
\begin{equation}
\label{eq:BC_MaxwellStress}
\llbracket \epsilon_0 \epsilon_\text{rel} \vec{E} \rrbracket  \cdot \vec{n} = q,
\end{equation} 
where $\llbracket A \rrbracket = A_1-A_2$ denotes the difference in $A$ between domain 1 and domain 2 and $\vec{n}$ the outward normal vector at the interface in domain 1. 
At the interface, charge conservation is expressed by a surface charge tansport equation \cite{Saville1997,Collins2008} as 
\begin{equation}
\label{eq:surface_charge_transport_eq}
\frac{\partial q}{\partial t} + \nabla_s \cdot \left( q \vec{u} \right) - D_s \nabla_s^2 q = \left\llbracket -K \vec{E} \right\rrbracket \cdot \vec{n}, 
\end{equation}  
where $\nabla_s$ denotes the gradient along the interface, $\vec{u}$ the fluid velocity, $D_s$ the surface diffusion coefficient, and $K$ the conductivity of the fluid. The term on the right hand side denotes the source due to ohmic conduction.
The normal electric stress on an interface results as 
\begin{equation}
\label{eq:Maxwell_Stress_interface_expression}
\left[ \sigma^M \cdot \vec{n} \right] \cdot \vec{n} 
= \frac{1}{2}\left\llbracket \epsilon_0 \epsilon_\text{rel} \left(\vec{E}\cdot \vec{n}\right)^2 
- \epsilon_0 \epsilon_\text{rel} \left(\vec{E}\cdot \vec{t}_1\right)^2 
-\epsilon_0 \epsilon_\text{rel} \left(\vec{E}\cdot \vec{t}_2\right)^2 \right\rrbracket,
\end{equation}
where $t_1$, $t_2$ denote the surface tangential vectors. At an interface between a perfect dielectric and a perfectly conducting liquid, the electric field is directed normal to the interface and vanishes inside the conducting liquid. The interface forms an equipotential surface with an induced surface charge screening off the electric field.
The Maxwell stress on the interface leads to a deformation, in which a balance between the Maxwell stress, the capillary forces and hydrostatic pressure is achieved. It is usually of conical shape, which is well-known and referred to as the Taylor cone.
This system has been studied extensively due to its simplicity and prototypic character in various contexts \cite{Taylor1964, Collins2008}.
However, for the same set of control parameters under which a Taylor cone forms, instead of a cone, a dimple in the liquid-liquid interface can be formed. That is, instead of being attracted by the pin-electrode, the interface is deflected away from the electrode. This points to a different mechanism than the Maxwell-stress induced interface deformation described above. 

We conducted a number of experiments using the square glass container shown in Fig. \ref{fig:Exp_setup}(a), while observing the interface with a slight incidence angle through the top-phase using the macro-objective. The voltage between the pin electrode and the aqueous KCl solution was controlled manually with the high-voltage sequencer from \textit{Labsmith}, and subsequently increased over time until an instability was triggered. When the interface became unstable, the camera was triggered with an end-trigger, stopping the recording and keeping the video before the trigger signal in memory. The results of these experiments are mainly intended to provide an overview of typical instability modes and interface configurations, rather than a basis for a quantitative analysis. 

\begin{figure}[t]
\centering
\includegraphics[]{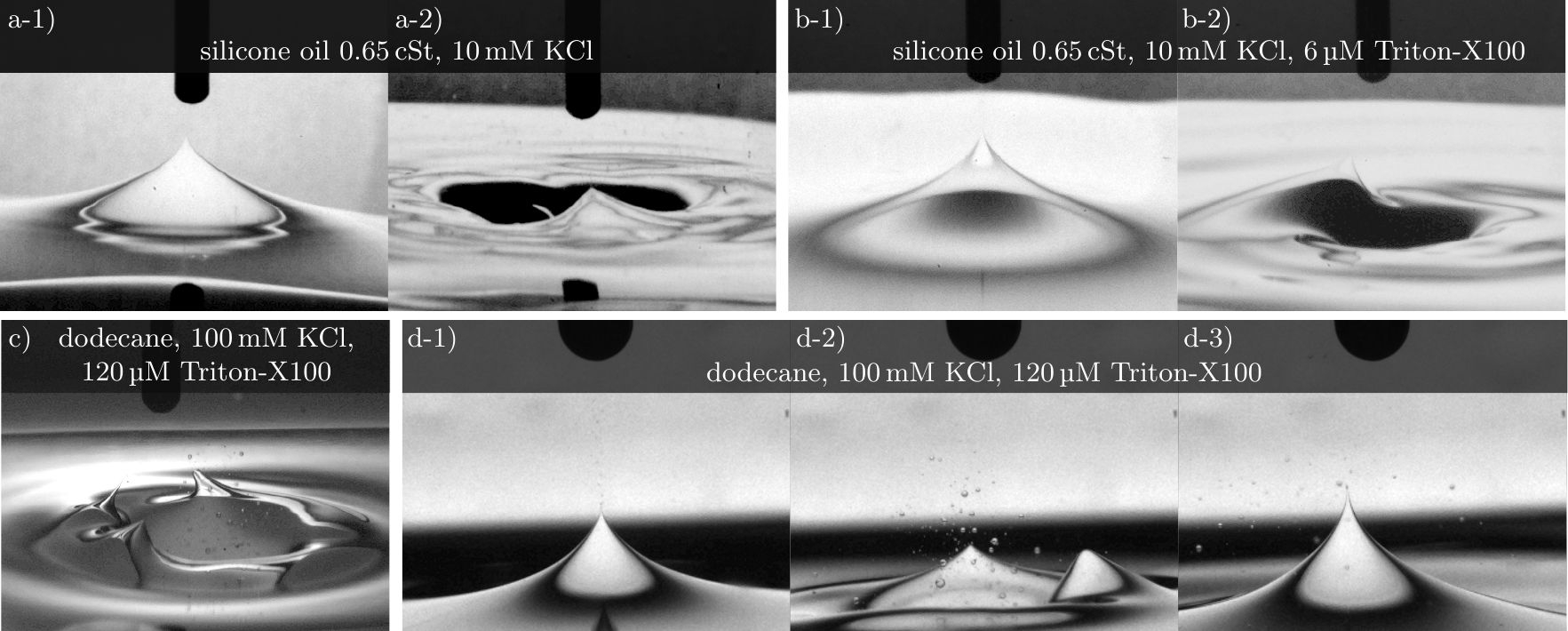}
\caption{\label{fig:EHD_instabilities} Exemplary deformation modes of liquid-liquid interfaces in a pin-interface configuration close to the time of surface disintegration. Images denoted with the same letter correspond to experiments conducted within the same system at similar experimental conditions. a) Interface between silicone oil ($\SI{0.65}{cSt}$) and a $\SI{10}{mM}$-KCl solution, forming a classical Taylor cone (1) or a dimple with cones emerging from its rim (2). b) Interface between silicone oil ($\SI{0.65}{cSt}$) and a $\SI{10}{mM}$-KCl solution with additional non-ionic surfactant (Triton-X100). Again, either a cone or a dimple is formed for similar applied voltages. c) Interface between dodecane and a KCl solution with a concentration of $\SI{100}{mM}$ and additional Triton-X100. A dimple forms with several cones emerging from the rim.  d) Image sequence showing the interface between dodecane and a KCl solution with a concentration of $\SI{100}{mM}$ and additional Triton-X100. Initially, the interface forms a classical Taylor cone (1), then transitions into a dimple (2), and subsequently forms a cone again (3). }
\end{figure}

Figure \ref{fig:EHD_instabilities} shows some representative surface deformation modes, with similar experimental conditions leading to strongly deviating results. Figure \ref{fig:EHD_instabilities}(a) shows a silicone oil with a viscosity of $\SI{0.65}{\centi St}$ above an aqueous KCl solution of $\SI{10}{\milli M}$, with an electrode-interface distance of $\SI{4}{\milli \meter}$. During the first experiment, the interface disintegrated at $\phi_n = \SI{5}{\kilo V}$ in the classical Taylor cone mode (Fig. \ref{fig:EHD_instabilities}(a-1)). During the final stages of the breakdown, droplets get ejected from the cone towards the electrode, and as soon as the system forms a conducting bridge, the power source shuts down. Upon repeating the experiment, a voltage of $\phi_n = \SI{4.5}{\kilo V}$ leads to a dimple at the interface (Fig. \ref{fig:EHD_instabilities}(a-2)), i.e., the interface moves away from the electrode.
From the rim of the dimple, a cone emerges and ejects droplets towards the electrode. Therefore, for very similar experimental conditions strongly different interface deformation modes are observed. 

Liquids can contain surface-active additives that change their properties and are responsible for a range of counter-intuitive results in free-surface flows, e.g. immobilized gas-liquid interfaces at superhydrophobic surfaces \cite{Peaudecerf2017}. In Fig. \ref{fig:EHD_instabilities}(b), results obtained with a similar system as the one of Fig. \ref{fig:EHD_instabilities}(a) are shown, where we added $\SI{5}{\micro \liter}$ of Triton-X100 (non-ionic surfactant, \textit{Sigma Aldrich}) to the lower phase, resulting in a surfactant concentration of $\SI{2.7}{\percent}$ cmc (critical micelle concentration). The experimental observations agree qualitatively with what was observed before: When the experiment is conducted for the first time, a cone that ejects droplets forms. When the experiment is repeated within the same system, the breakdown mechanism changes to a dimple at the interface, while a cone emerges from the rim. Thus, it is unlikely that surfactants play a role for the observed phenomena.

The formation of dimples at the interface can be demonstrated for a wide variety of dielectric liquids and KCl concentrations, where for similar experimental conditions different interface deformation modes are observed. The values of the critical voltages vary, as well as the nature of the breakdown mechanism. For example, in Fig. \ref{fig:EHD_instabilities}(c), we show results obtained with dodecane (\textit{Sigma-Aldrich}) as the dielectric liquid, with higher concentrations of both KCl ($\SI{100}{\milli M}$) as well as Triton-X100 ($\SI{100}{\micro\liter}$) in the aqueous phase. Interestingly, the dimple still forms, but now several cones emerge simultaneously from the rim. A wide range of phenomena can be observed, but in the present context we are not interested in creating a map of the different dynamic regimes. Instead, we focus on the mechanisms responsible for the different interface deformation modes, especially the emergence of a dimple. 

Figure \ref{fig:EHD_instabilities}(d) shows a series of three images from the same experiment with dodecane as the dielectric liquid and a KCl- and Triton-X100 solution as the lower phase. Initially, at a voltage difference of $\SI{4920}{V}$ and an electrode spacing of $\SI{5}{\milli \meter}$, a Taylor cone forms and ejects droplets towards the electrode. The droplets move towards the electrode, and change direction after contact with the electrode. Then, they move towards the interface, and shortly thereafter, the liquid-liquid interface displays a dimple with cones emerging from the rim. Subsequently, the cone at the center forms again. The appearance and disappearance of the dimple strongly correlates with the existence of droplets in the dielectric liquid. When droplets are present, a dimple forms, and when the droplets have merged with the main phase, the cone at the center is observed. Revisiting the other experiments, droplets are visible in (c), whereas in (a) and (b) the droplets are not immediately visible. It can be hypothesized, however, that during the first experiment some electrolyte droplets were deposited at the pin electrode. Upon repeating the experiment with the same system, droplets would get emitted from the upper electrode. It is likely that the size of these droplets is below the resolution limit of the macro-objective. Such a scenario would point to strong memory effects of the system: The interface deformation mode would depend on past evolution of the system. 

To test the hypothesis that droplets play a crucial role for the interface deformation modes, it is necessary to understand how exactly droplets influence the system. In that context it is important to note that while other factors such as surfactants might not be the original cause for the different surface deformation modes, they may change the way how droplets influence the system. For example, the surfactant concentration reduces the interfacial tension, which in turn influences the production of droplets and weakens the resistance to dimple formation. Several mechanisms could be responsible for the observed changes in the interface deformation modes: \\
i) \textit{Inverted Maxwell stress}: 
The effects of the charged droplets can be interpreted as an increase of the conductivity of the oil phase (domain 1). In the case of a steady-state system with dominant conduction, eq. \ref{eq:surface_charge_transport_eq} results in a ratio of the normal electric field components as $\vec{E}_1\cdot \vec{n} = K_2/K_1 \vec{E}_2 \cdot \vec{n}$. The Maxwell stress at an interface can be expressed according to eq. \ref{eq:Maxwell_Stress_interface_expression} as 
\begin{equation}
\label{eq:Maxwell_stress_inverted_hypothesis}
[ \sigma^M \cdot \vec{n} ] \cdot \vec{n} 
= \frac{1}{2} \epsilon_0  
\left( 
\epsilon_\text{rel,1} -   \epsilon_\text{rel,2} \left( \frac{K_1}{K_2} \right)^2 
\right) 
\left( \vec{E}_1 \cdot \vec{n} \right)^2. 
\end{equation}
The droplets in the oil phase transport charge from the upper electrode to the interface, and with sufficiently large conductivity in the oil phase, the sign of the Maxwell stress could change relative to a system without conduction. As a result, the electrode would repel the liquid-liquid interface. \\
ii) \textit{Droplet impact}: The droplets impacting the liquid-liquid interface carry momentum, which could lead to a deformation of the interface upon impact. If the momentum of the impacting droplets is large enough, it could overcome the Maxwell stress at the interface.\\
iii) \textit{Viscous momentum transfer}: While traveling through a liquid, droplets transfer momentum to the surrounding liquid. If the viscous momentum transfer is large enough, the dielectric liquid could develop a background flow that interacts with the interface. This, in turn, could cause the formation of a dimple. \\
All of the above hypotheses rely on the existence of electrolyte droplets within the dielectric liquid to explain the observed dimple. 
In the pin electrode system, the droplet ejection process is hardly controllable, as it occurs when the interface breaks down. Subsequent processes such as droplets exchanging charge with the electrode, secondary breakup and coalescence, and highly dynamic interface deflections additionally limit our ability to access the relevant mechanism. Therefore, we replaced the pin electrode with a hollow metallic needle, as shown in Fig. \ref{fig:Exp_setup}(b). Thus, by utilizing a syringe pump we can control the amount of electrolyte phase introduced into the dielectric liquid at the upper electrode via the flow rate. Additionally, the occurrence of droplets is decoupled from the interface breakdown, as we can introduce droplets at a subcritical voltage before a Taylor cone forms.
The droplets emitted from the metallic needle travel through the dielectric fluid towards the liquid-liquid interface, and merge with the electrolyte solution. In case of the pin-interface configuration, the droplets move towards the electrode and invert their charge, which has been described previously \cite{Mochizuki1990, Hase2006,Jung2008}.
Next, they travel towards the liquid-liquid interface, where they merge with the electrolyte solution.
As a result, we are able to mimic the behavior of the droplets in the pin-interface configuration using a submerged electrospray.
Therefore, we are able to investigate the cause of the interface deflection under well-controlled conditions. 

\section{Quantification of interface deflection}
\label{sec:needle_interface_configuraion}

\begin{figure}[tb]
\centering
\includegraphics[]{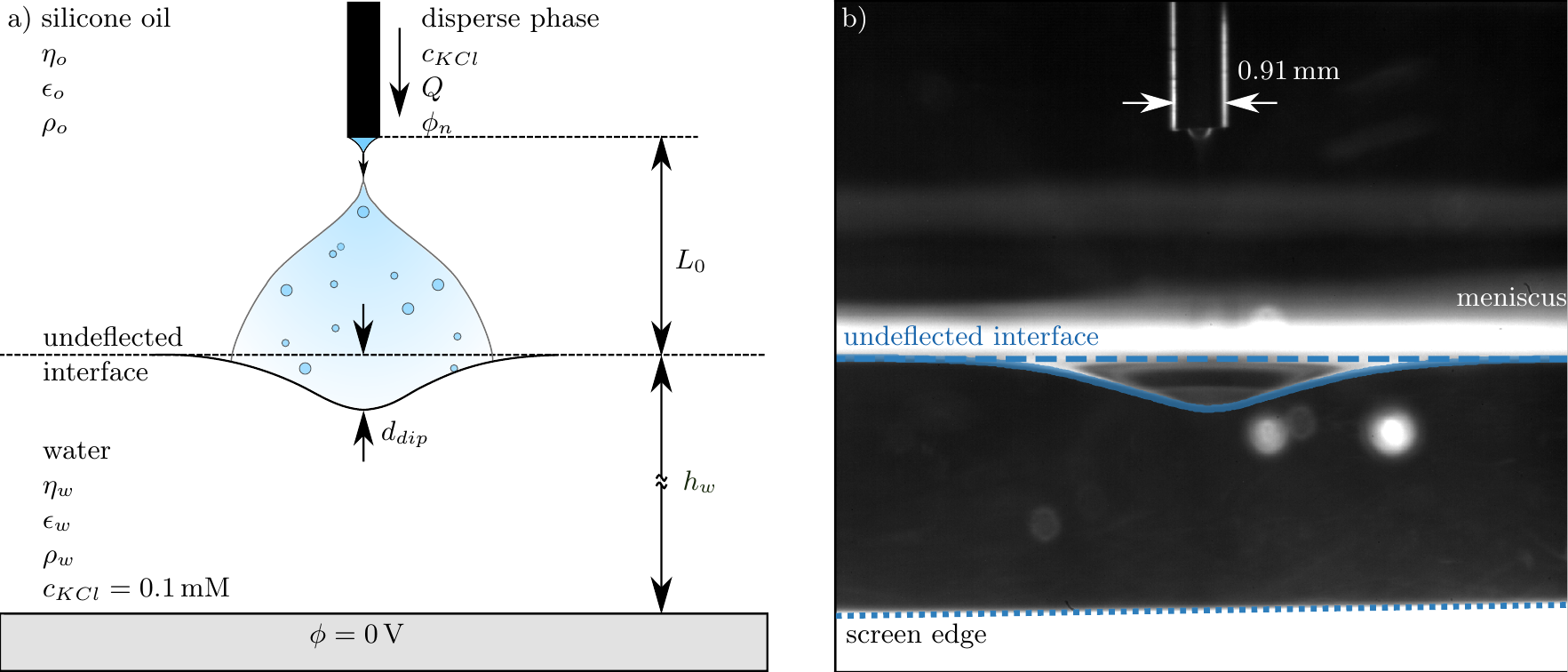}
\caption{\label{fig:Dip_observation} Characterization of the interface deflection. a) Schematic of the system, including the relevant parameters. The upper dielectric phase consists of a silicone oils of different viscosities $\eta_o$, dielectric permittivity $\epsilon_o$ and mass density $\rho_o$. The lower phase is an aqueous KCl solution with a concentration of $c_{KCl} = \SI{0.1}{\milli M}$. The lower electrode in contact with the conducting liquid is held at a constant potential of $\SI{0}{V}$, whereas the potential $\phi_n$ at the metallic needle is varied. In addition, the KCl concentration $c_{KCl}$ of the emitted liquid at the needle is varied, as well as the volume flow $Q$. The droplets emitted from the needle are dispersed in the oil phase as an electrospray (blue region) and merge with the lower phase. 
b) Representative interface deflection observed in sideview. The average interface position (solid blue line) is extracted from an average image that is created from the recorded video (see main text for details). Above the undeflected interface (dashed blue line), the optical distortion caused by the meniscus at the cylinder side wall is visible as a bright region. At the bottom, the boundary of the cardbox matte screen is marked by the dotted line, illustrating that by imaging the reflection of the interface the contrast between interface and background is well visible. 
}
\end{figure}

The droplets ejected from the needle lead to an interface deflection, which we characterize for varying experimental parameters. We performed the experiments described in the following using the circular cell illustrated in Fig. \ref{fig:Exp_setup}(b) and the \textit{Heinzinger} high voltage source. Fig. \ref{fig:Dip_observation}(a) shows some key parameters of the experimental setup. Both the needle diameter ($D_n=\SI{0.91}{\milli \meter}$) and the needle-interface distance $L_0=\SI{5}{\milli \meter}$ are constant. Silicone oils of varying viscosities ($\SI{0.65}{\centi St}$ (\textit{Wacker AK 0.65}), $\SI{1}{\centi St}$, $\SI{5}{\centi St}$ and $\SI{10}{\centi St}$ (\textit{Elbesil B})) serve as dielectric liquid. Detailed liquid properties can be found in Appendix \ref{sec:Supp_Liquids}. The lower phase consists of DI-water (\textit{Milli-Q}) with a KCl concentration of $c_{KCl} = \SI{0.1}{\milli M}$ and a constant filling height of $h_w=\SI{20}{\milli \meter}$. By holding these properties constant, we ensure that the electric conduction timescale of the lower phase remains constant. Also, due to the comparatively large container size, the emitted electrolyte does not change the distance $L_0$, since the emitted volume is comparatively small. 

In order to change the electrospray properties, we varied the flow rate and the conductivity of the electrolyte during the parameter studies.
The operation parameters were chosen such that the interface deflection was stable, and we determined their range in preliminary studies. 
We varied the applied potential difference between \SI{2000}{\volt} and \SI{4000}{\volt}, since this represents the voltage range slightly below the onset of EHD tip streaming in the pin-interface configuration.
The minimum flow rate for which the interface depression was stable was identified as \SI{20}{\micro\liter\per\minute}, and for the purpose of studying the influence of the flow rate, $Q$ was increased up to \SI{80}{\micro\liter\per\minute}.
Below a flow rate of \SI{20}{\micro\liter\per\minute}, the interface deformation became unstable, which can be rationalized by the electrospray emission mode from the needle. For all configurations, the emission of droplets was instationary, with oscillations at the meniscus leading to oscillating droplet ejections. As a result, the forcing mechanism of the interface is expected to exhibit a similarly unsteady behavior, leading to transient deformations. At higher flow rates, the emission remained unstable, while more dispersed liquid volume was present in the system.
Then, the average time between successive droplet emissions becomes sufficiently small compared to the relaxation time of the interface, resulting in a forcing of the interface that is sufficiently homogeneous to observe a stationary interface deformation.
We chose the conductivity of the electrolyte to be relatively high, in the range between \SI{10}{\milli\siemens\per\meter} and \SI{1.65}{\siemens\per\meter}. 

We would like to emphasize that the parameters were not chosen to achieve a specific electrospray regime, e.g., the stable cone-jet mode, but rather to fix the interface deformation mode.
Nonetheless, we can compare our configuration to the regimes described by Higuera \cite{Higuera2010} for submerged electrosprays.
In principle, the scaling described by Higuera is only applicable to steady cone-jets, but it provides some insight into the overall spray regime. 
As discussed by \citeauthor{Higuera2010}, the viscosity of the dielectric liquid is of importance compared to inertia and axial viscous forces of the inner liquid if the dimensionless parameters $\Pi = \mu_o K^{1/3} / (\rho_w \epsilon_o \gamma^2)^{1/3}$ and $\Pi R^{1/4}$ with $R= \rho_w^{4/3} K^{2/3}  Q / (\epsilon_o^{2/3}  \mu_w  \gamma^{1/3})$ are of the order of unity or larger.
Here, $\mu_o$ and $\mu_w$ are the dynamic viscosities of the dielectric liquid and the aqueous phase, respectively, $\rho_w$ is the density of the aqueous phase, $\epsilon_o$ the relative permittivity of the dielectric, $\gamma$ the interfacial tension and $K$ the conductivity of the disperse phase.
For the specific case of the silicone oil with \SI{1}{\centi St} viscosity, a spray flow rate of \SI{20}{\micro\liter\per\minute} and a KCl concentration of \SI{1}{\milli\mol\per\liter}, the values are $\Pi = 1.26$ and $\Pi R^{1/4} = 16.1$, indicating that the outer liquid viscosity has a strong effect on the spray.
Higuera provided a second estimate with respect to the flow rate. 
If the flow rate is of the order $Q_m = \epsilon_o \gamma^2 D_i  /(3 \mu_o^2 K)$ or larger, and if the viscosity of the outer liquid is important, no stable jet configuration can exist.
For the given parameters, we obtain $Q_m =\SI{7.9}{\micro\liter\per\minute}<Q$.
Thus, the liquid stream is expected to suffer from instabilities in close proximity to the needle, breaking up into droplets.
Also, the current is expected to scale approximately as $I \propto \epsilon_o^{1/2} \gamma^{3/2} (D_i/2)^{1/2} / \mu_o$, where $D_i$ denotes the inner needle diameter, independent of $Q$.
During the experiments, we observed that the Taylor cone at the needle was oscillatory, and that instabilities occurred close to the needle, being in qualitative agreement with the scaling relations of \citeauthor{Higuera2010}.
To summarize, in the regime of high flow rates (compared to the steady cone-jet mode), the electrospray is expected to be influenced by the viscosity of the dielectric liquid.

The system response can be characterized by two observables, the transported current $I$ and the interface deflection $d_\text{dip}$. In order to quantify how the interface deflection changes, we choose parameters that lead to a steady-state dimple. Therefore, we omit transient effects, simplifying the subsequent evaluation. During each experiment, the flow rate $Q$, the disperse phase KCl concentration $c_{KCl}$ and the applied voltage $\phi_n$ are fixed. The ammeter monitors the current. We utilize the ammeter's built-in averaging function (moving average with 100 readings, with a measurement period of $1/\SI{60}{s}$) to minimize measurement uncertainties. After applying the voltage, we wait until the averaged current stabilizes. Then, we record a video with the high-speed camera at $1000$ frames per second ($\SI{}{fps}$) with a resolution of $1024\times 1024$ pixels ($\SI{}{px}$), comprising a total of 200 images.

We extract an average image from the recorded video, which we use in turn to determine the interface deflection. First, utilizing \textsc{python 3.7} we apply a moving minimum filter on the video, where we average 12 images along the time-axis. This step ensures that short-term fluctuations such as droplets are filtered out from the image, and only the reflections from the interface remain. In a second step, we average the 200 frames and extract one image by computing the mean greyscale value of each pixel. Figure \ref{fig:Dip_observation}(b) shows a representative result. Here, the interface is clearly visible due to the reflections from the LED panel, whereas the background appears dark due to the black screen, leading to a large signal-to-noise ratio. At the bottom, we have marked the edge of the screen. Also, the side-wall meniscus is visible as a bright area just above the deflected interface. From the average image, we extract the interface using \textsc{python 3.7}. First, we use an edge-detection mechanism to identify all potential interfaces. Then, a starting point on top of the interface is determined by user input, and all interconnected edges are identified. In Fig. \ref{fig:Dip_observation}(b), the blue line indicates the identified interface, from which the maximum deflection $d_\text{dip}$ is computed. To extract the height in physical coordinates, we calibrate the optical system using an image of a reference grid. 

\subsection{Effect of the flow rate}

\begin{figure}[tb]
	        \centering{
			\includegraphics[]{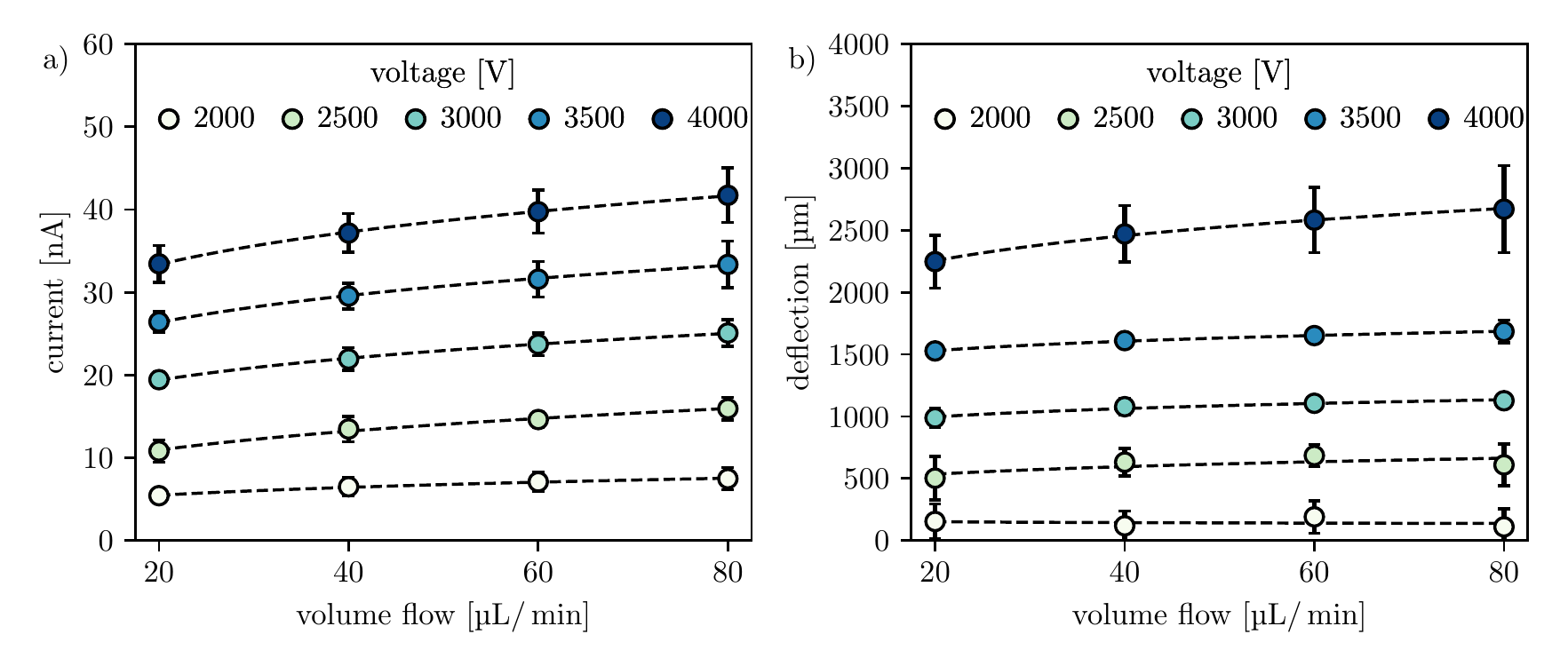}
	        \caption{Measured current (a) and deflection (b) as a function of the volume flow $Q$ and the applied potential difference $\phi$. The silicone oil viscosity was $\SI{5}{cSt}$ and the KCl concentration of the spray $\SI{1}{mM}$. Error bars represent the standard deviation of 5 measurements per datapoint. The dashed lines represent fit functions of the form $y = a x^b$ to guide the eye.
}        
	        \label{fig:Voltages_volumeflow_5cSt}
	        }
\end{figure}

Figure \ref{fig:Voltages_volumeflow_5cSt} shows the experimental results for a silicone oil with a viscosity of $\SI{5}{cSt}$ and a KCl concentration of the disperse phase of $\SI{1}{\milli M}$, with volume flows varying in the range $Q=20-\SI{80}{\micro \liter\per \minute}$ and varying applied potential differences $\phi$. As Fig. \ref{fig:Voltages_volumeflow_5cSt}(a) shows, the current increases roughly as $I \propto Q^{0.2}$, exhibiting a weak dependence on the volume flow. A similar dependence was reported for a microscopic co-flowing electrospray by Gundabala \textit{et al.} \cite{Gundabala2010}. Figure \ref{fig:Voltages_volumeflow_5cSt}(b) shows the interface deflection, exhibiting only a weak dependence on volume flow as well. By contrast, the voltage $\phi$ has a strong influence on both the current and the deflection. The current increases approximately linearly with the voltage, similarly to the results reported by Gundabala and coworkers. The deflection shows an even stronger dependence on the voltage. Since the dependence on volume flow is weak, we will keep the volume flow constant for the rest of the parameter studies, with a fixed value of $Q = \SI{40}{\micro \liter \per \minute}$.

\subsection{Effect of the viscosity}
\begin{figure}[t]
	        \centering{
			\includegraphics[]{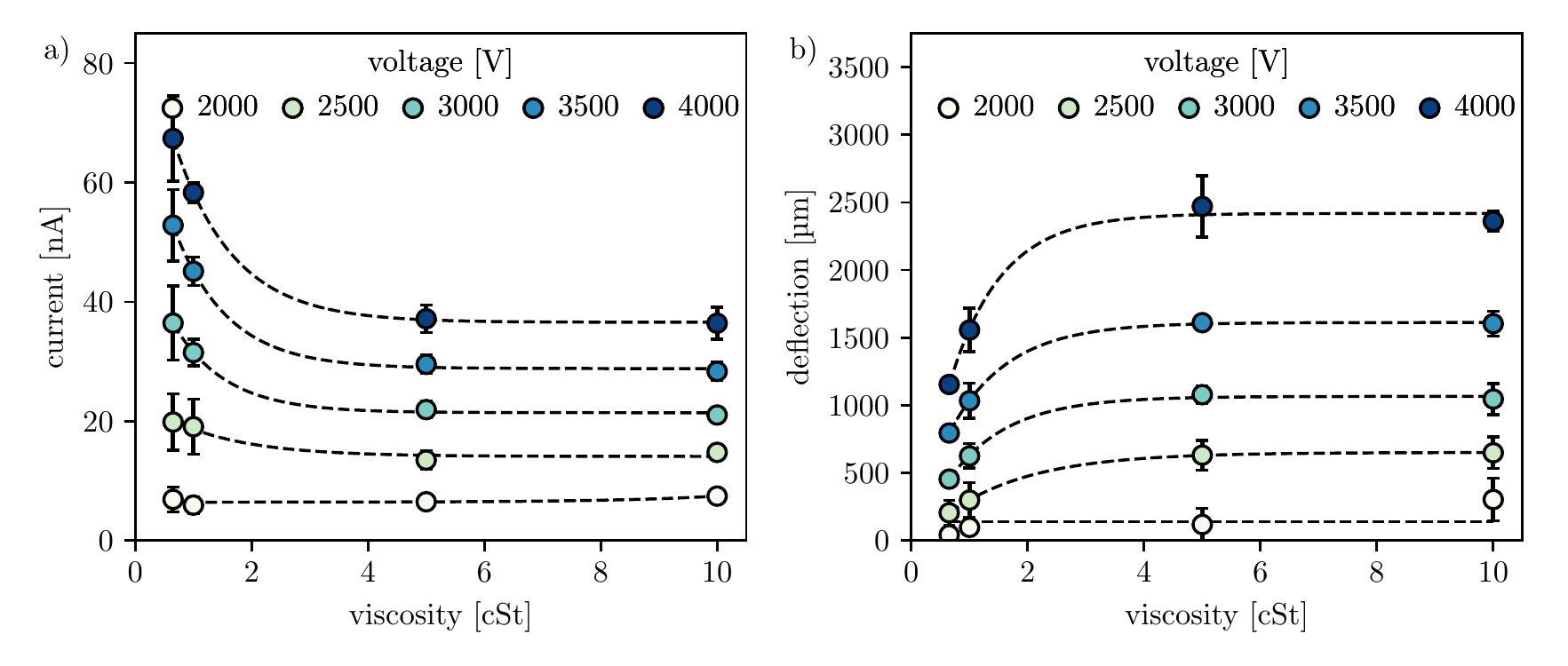}
	        \caption{Measured current (a) and deflection (b) as a function of the silicone oil viscosity and the applied voltage $\phi$. The volume flow was $\SI{40}{\micro \liter \per \minute}$ and the KCl concentration of the spray $\SI{1}{mM}$. Error bars represent the standard deviation of 5 measurements per datapoint. The dashed lines represent exponential fit functions to guide the eye.
	        }        
	        \label{fig:viscosity}
	        }
\end{figure}

Figure \ref{fig:viscosity} shows the influence of the viscosity of the dielectric liquid on the current as well as the deflection, with the experimental parameters fixed at $Q=\SI{40}{\micro \liter\per \minute}$ and  $c_{KCl} = \SI{1}{\milli M}$. As can be seen from Fig. \ref{fig:viscosity}(a), the current decreases with increasing viscosity. However, while the viscosity change from $\SI{1}{\centi St}$ to $\SI{5}{\centi St}$ leads to a current decrease of the order of $\SI{30}{\percent}$, a further increase from 
$\SI{5}{\centi St}$ to $\SI{10}{\centi St}$ has no significant effect. This behavior indicates a limiting mechanism becoming important at higher viscosities. While the current decreases, the deflection increases with viscosity. For the highest voltage, the deflection increases from about $\SI{1.2}{\milli \meter}$ at $\SI{0.65}{\centi St}$ to $\SI{2.5}{\milli \meter}$ at $\SI{5}{\centi St}$. Again, a further increase of viscosity has no further significant effect. 

These results indicate that the deformation is not due to an \textit{inverted Maxwell stress} induced by the charge transport of the droplets. As we have noted, the ion concentration of the lower liquid layer is held constant at $c_{KCl} = \SI{0.1}{\milli M}$, and thus its conductivity remains constant as well. 
An increasing current is equivalent to a higher effective conductivity $K_1$ within the oil layer, and following eq. \ref{eq:Maxwell_stress_inverted_hypothesis}, the interface deformation should increase. However, since the deflection increases with decreasing current, it is reasonable to assume that this mechanism is not responsible for the interface deflection. 

\subsection{Effect of KCl concentration}
\label{subsec:KCl_Concentration}
\begin{figure}[t]
	        \centering{
	        \includegraphics[]{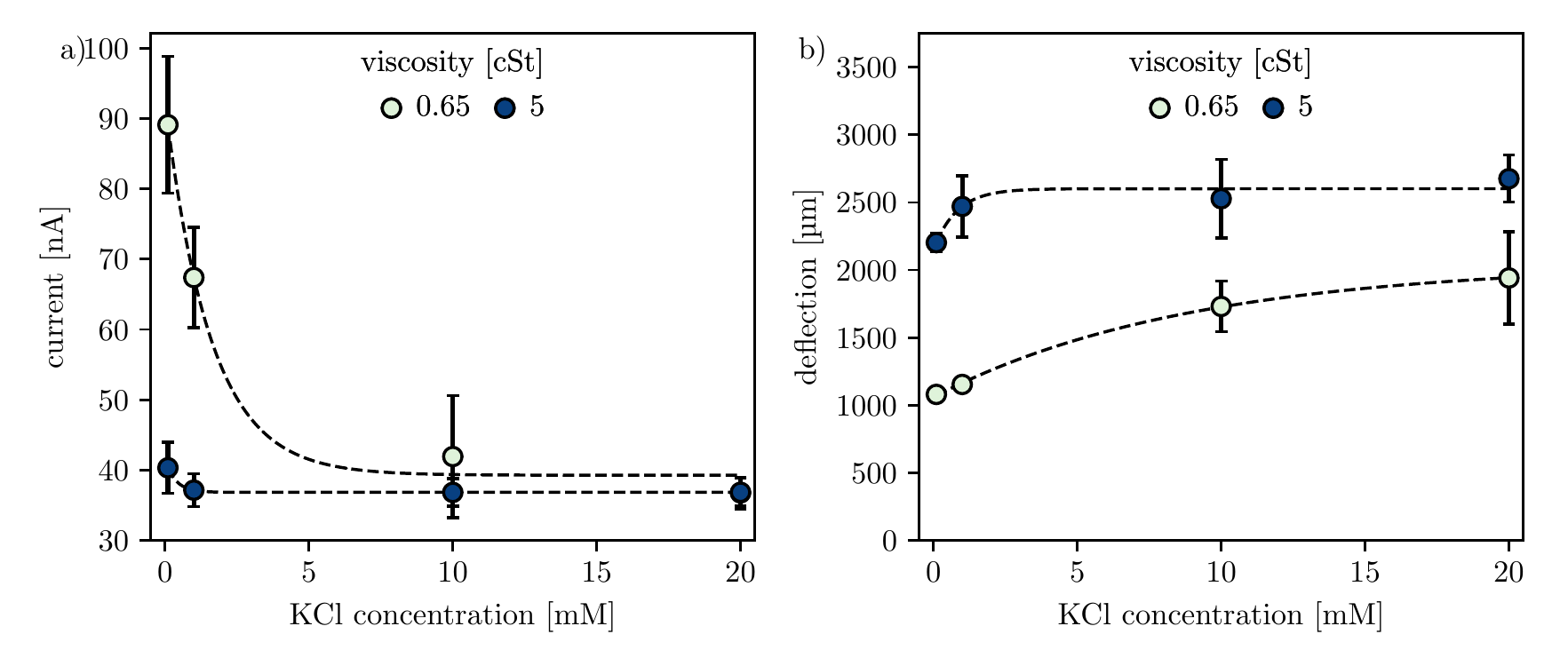}
	        \caption{Measured current (a) and interface deflection (b) as a function of the KCl concentration and the oil viscosity for a fixed volume flow of $\SI{40}{\micro \liter \per \minute}$ and a voltage of $\SI{4000}{V}$. Error bars represent the standard deviation of 5 measurements per data point. The dashed lines represent exponential fit functions to guide the eye.
	        }        
	        \label{fig:concentration}
	        }
\end{figure}

The KCl concentration determines the conductivity $K$ of the electrolyte, leading to a higher conductivity at higher ionic strengths. As shown in Appendix \ref{sec:Supp_KCl_cond}, the conductivity $K$ is nearly linearly dependent on the ionic concentration in the considered concentration range. Figure \ref{fig:concentration} shows the resulting current and interface deflection for KCl concentrations of the sprayed liquid varying between $\SI{0.1}{\milli M}$ and $\SI{20}{\milli M}$, for a constant applied voltage of $\SI{4000}{V}$ and a fixed volume flow of $\SI{40}{\micro \liter \per \minute}$. Two silicone oil viscosities are shown, $\SI{0.65}{\centi St}$ and $\SI{5}{\centi St}$. For $\SI{0.65}{\centi St}$, the current depends strongly on the KCl concentration and decreases with higher concentrations, whereas the current for $\SI{5}{\centi St}$ is nearly constant over the range of the concentrations. With increasing KCl concentration, the current approaches the value of the current for $\SI{5}{\centi St}$. Especially the result obtained for $\SI{0.65}{\centi St}$ is surprising, because results reported in literature show an inverse behavior. For example, during steady cone-jetting, a proportionality  $I\propto K^{0.5}$ was reported, initially observed in air, but was also shown to hold for a range of submerged electrosprays \cite{DelaMora2007}. In a work by Mar\'{i}n \textit{et al.} \cite{Marin2012}, surface tension effects were analyzed in a submerged electrospray, with droplets created in a hexane bath. The observed current followed the classical $I \propto K^{0.5}$ dependency for a wide range of parameters. However, it is important to note that in the referenced work, liquids with lower conductivities were used, and that the flow rates were smaller than the ones we applied, leading to steady cone-jet regimes. 

The deflection of the interface as shown in Fig. \ref{fig:concentration}(b) increases with increasing KCl concentration for $\SI{0.65}{\centi St}$, while it remains nearly unaffected for $\SI{5}{\centi St}$. This behavior is inverse to that of the observed current, and a similar limiting behavior is found for high concentrations. As already found above, the deflection is not proportional to the current, thus giving further evidence that the \textit{inverted Maxwell stress} hypothesis has to be discarded. 


\subsection{Discussion}

The experimental results give an indication about which mechanisms presumably cause the surface depression.
In order to resolve the contradiction between the expected and the observed current as a function of KCl concentration, it is instructive to clarify how the KCl concentration affects the spray droplet size.
The disintegration of the liquid-liquid interface under an electric field is a balance between competing mechanisms. 
During steady cone-jetting, a higher conductivity leads to smaller jet diameters, which in turn lead to smaller droplets during successive jet break up. 
The exact diameter and breakup mode of the jet depend on how mechanical stresses are balanced (see e.g. reference \cite{Higuera2010}). 
Usually, however, a higher conductivity leads to a smaller radius. 
In our experiments, the jet emission is unstable and highly dynamic, thus these observations cannot be translated directly. 
In addition to the initial jet breakup, secondary droplet breakup leads to smaller droplets. 
Droplets with a charge larger than a critical charge are unstable and emit smaller droplets. This critical charge is called the Rayleigh charge and is given as 
\begin{equation}
\label{eq:Rayleigh_charge}
q_R = 8 \pi \sqrt{\epsilon _0 \gamma R^3},
\end{equation}
where $\gamma$ represents the interfacial tension and $R$ the initial droplet radius \cite{Rayleigh1882}. Experimental results by Hunter and coworkers \cite{Hunter2009} show that the radius of the emitted progeny droplets in air scales as $R_d \propto K^{-2/3}$. In addition to the size, the charge of the progeny droplets $q$ is affected by the concentration as well. For example, Hunter \textit{et al.} observed a constant ratio of the progeny drop charge and the Rayleigh charge ($q/q_R \approx \mathit{const.}$), which leads to a proportionality of $q\propto R_d^{3/2}$. Pillai and coworkers found a different proportionality as $q/q_R \propto (R_d/R)^{5/2}$, where $R$ denotes the radius of the mother droplet and $R_d$ the radius of the progeny droplet. However, their results were obtained for microscopic droplets of initial sizes in the submicron range and might not be representative for larger droplets. As we have noted before, the determination of the exact droplet size and charge distribution is challenging, making us reliant on these previous results. Additionally, electrosprays operated above the minimal flow rate of the steady cone-jet have been reported to exhibit non-uniform droplet size distributions (e.g. references \cite{Marin2012,Rosell-Llompart1994}). Especially for unsteady droplet emissions, satellite droplets with sizes of a few $\si{\micro m}$ form, with a constant size for varying operating conditions \cite{Marin2012}. Thus, we expect a similar effect in our system, leading to the formation of satellite droplets. 

As a first approximation, we can compute the droplet velocity by treating them as spheres with an immobilized interface. Then, based on the Stokes drag force \cite{Stokes1851} and by balancing electric forces and drag forces on the droplet, the terminal velocity is given by 
\begin{equation}
\label{eq:Stokes_velocity}
u_d = \frac{q_d E}{6 \pi \mu R_d} \propto \frac{q_R E}{\mu R_d} \propto \frac{E R_d^{1/2}}{\mu} \propto  \frac{ E}{ K^{1/3} \mu},
\end{equation}
where we have assumed a constant ratio between the droplet charge and the Rayleigh charge as reported by Hunter \textit{et al.} and $R_d \propto K^{-2/3}$. While we could use other models such as Hadamard – Rybczynski equation to relax the condition of an immobilized liquid-liquid interface, the scaling argument persists. This result indicates that droplets will have lower velocities relative to the ambient fluid as the conductivity increases. Especially the small satellite droplets will remain longer in the dielectric phase. Since the droplets carry charge, it is reasonable to assume that they form an effective space charge region between the pin electrode and the interface. In turn, the space charge partially screens off the electric field, leading to a reduction of the emitted current. Interestingly, equation \ref{eq:Stokes_velocity} also provides a qualitative explanation for the limiting mechanism observed in the preceding section in context with the viscosity variations: An increase in viscosity leads to a decreased droplet velocity, resulting in a stronger accumulation of the space charge. This indicates that the self-regulatory behavior has the same origin for both the concentration as well as the viscosity. In section \ref{sec:numerical_modeling}, we will further investigate this limiting effect and demonstrate the space-charge regulation of the electric field at the emitting needle. 

The experimental results indicate that the relevant driving mechanism for the interface deflection is the \textit{viscous momentum transfer}. The background fluid is dragged along with the droplets. Since the magnitude of the drag force of the dielectric liquid on the droplets is the same as the magnitude of the drag force from the droplets on the dielectric, the effect of the droplets can be interpreted as an electric volume force density on the liquid given by
\begin{equation}
\label{eq:volume_force}
\vec{f}_\text{el}=\rho_\text{el} \vec{E},
\end{equation}
where $\rho_{el}$ is the electric space charge due to the droplets. The dielectric liquid is set in motion and a flow is generated that impinges on the interface, causing its deflection. The hypothesis is consistent with the observed increase of deflection with ionic concentration as well as viscosity, as fast moving droplets result in a smaller space charge than slowly moving droplets. In addition, it provides a self-regulating mechanism accounting for the plateauing at high viscosities and concentrations (c.f. Fig. \ref{fig:concentration}), as well as the decreasing current with increasing conductivity. 
Viscous momentum transport of an electrospray to a surrounding fluid was also observed for electrosprays in gaseous environments. Tang and Gomez measured the gas velocity in the proximity of an electrospray, and observed a significant flow velocity \cite{Tang1994} even for the small viscosity of air.
In the system under investigation here, the viscosity of the surrounding liquid is larger, which leads to smaller velocities of the droplets. As the droplets exhibit larger residence times, the total viscous momentum transfer of the electrospray due to electrostatic forces is expected to be larger.

The charge transport by the droplets is due to two mechanisms: 
First, the droplets have a relative velocity to the surrounding dielectric fluid, following equation \ref{eq:Stokes_velocity}. 
Second, the droplets are transported by the advective velocity of the surrounding medium.
Since the electrospray is operated at relatively high flow rates, it probably has a rather broad droplet size distribution. 
As the velocity $\vec{u}_d$ decreases with decreasing droplet radius, it is plausible that the motion of the small droplets is dominated by the background flow velocity. In that context, it could happen that droplets recirculate in the dielectric liquid without merging with the lower phase.
This would imply that a fraction of the current emitted from the electrospray does not reach the lower electrode.
In order to judge whether or not such an entrainment effect is significant, additional measurements (not shown) for the highest viscosity and the highest electrolyte conductivity were performed at varying applied voltages, where either the emitted current at the needle or the current at the lower electrode were measured. 
The measurements showed no significant differences between the two current values, indicating that entrainment by the dielectric fluid does not significantly affect the measured current. 
A likely reason is that the total current associated with the smaller droplets is small, which was for example shown by Tang and Gomez for electrosprays in air \cite{Tang1994}.

\section{Numerical modeling}
\label{sec:numerical_modeling}

In order to verify that the impinging background flow results in the interface deflection, we reproduce the observed background flow in the dielectric liquid using an effective model. We use \textsc{Comsol Multiphysics~5.5}, a finite element method suite, and compare the results to our experimental observations. 
The underlying idea is to self-consistently determine the background velocity of the dielectric liquid, incorporating fluid flow, electrostatics and the charge transport by the droplets.
To compensate for potential modeling inaccuracies, we fix the free parameters of the model based on particle tracking measurements.
As we want to isolate the effect the flow in the dielectric liquid has on the liquid-liquid interface, we do not account for the Maxwell stress at the interface.
This simplification is supported by the fact that in the absence of an imposed flow rate $Q$, we could not observe a significant interface deformation for the electrospray parameters under investigation.

\begin{figure}[tb]
\centering
\includegraphics[]{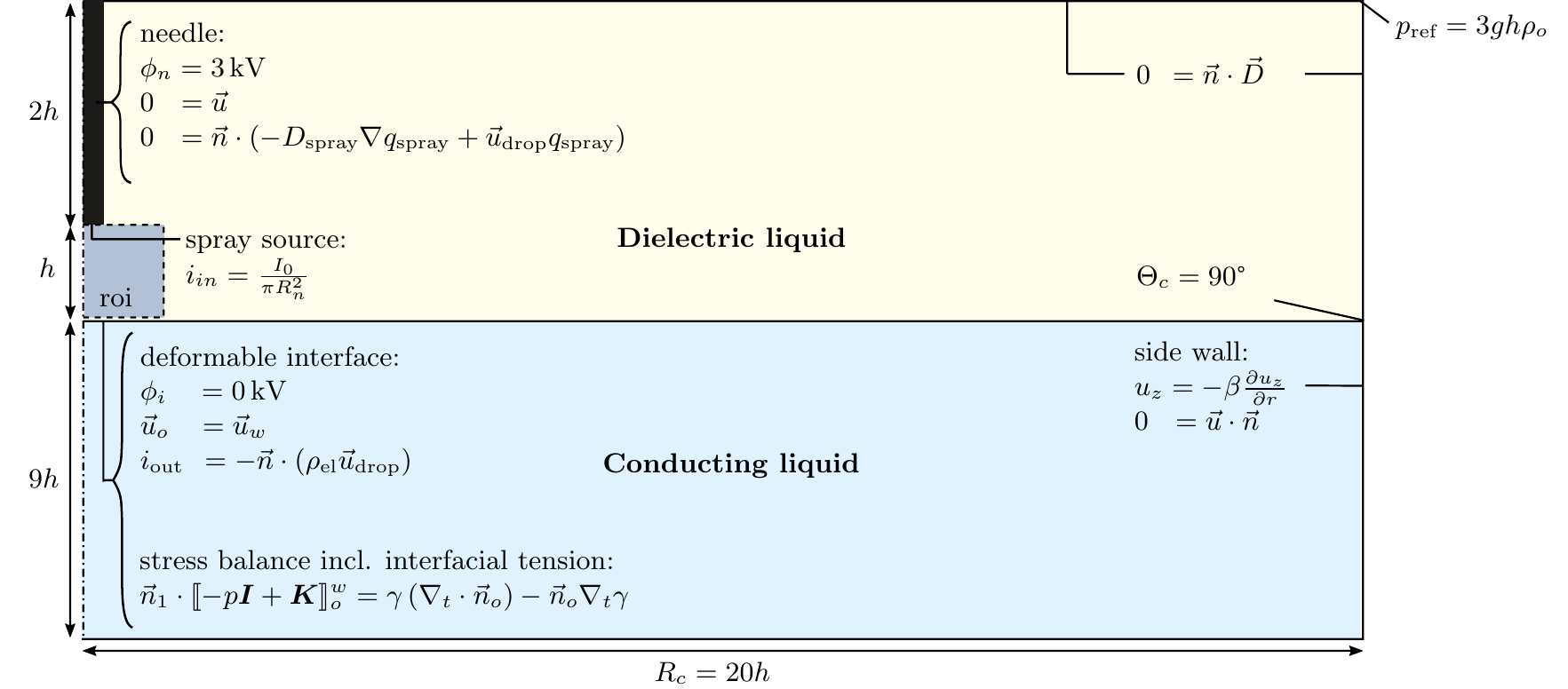}
\caption{\label{fig:numerics} Schematic of the numerical model with boundary conditions as implemented in \textsc{Comsol Multiphysics 5.5}. The axial symmetry of the problem is utilized, and the liquid-liquid interface can deform in reaction to the normal stress.
The model comprises the electrostatic problem, fluid flow and charge transport within the dielectric liquid, and fluid flow in the lower phase.}
\end{figure}

Figure \ref{fig:numerics} shows a schematic of the computational domain with the relevant boundary conditions. We utilize the axial symmetry of the problem and set the center line as a symmetry axis.
In order to simplify the numerical treatment, we model the charge transport from the needle to the interface by an advection-diffusion equation.
Here, the charge transported by the spray is treated in an Euler-Euler framework of two interpenetrating media.
The underlying idea is that the current is transported through the dielectric phase by droplets only, where the droplet velocity is determined by the advective velocity due to the background dielectric and the relative electrophoretic motion of the droplets.
By modeling the charge transport, we are able to compute the charge distribution inside the dielectric phase, which in turn defines the electric force density entering the Navier-Stokes equations. 
As a simplifying assumption, we model the droplets as monodisperse with radius $R_d$, which allows us to compute the electrophoretic velocity of the droplets.
As we have discussed in section \ref{subsec:KCl_Concentration}, the literature on the droplet charge and size distribution is sparse, and we have to introduce assumptions about their distribution. 
We model the droplet charge as a given fraction of the Rayleigh charge  $q_d/q_R = 0.44$ (eq. \ref{eq:Rayleigh_charge}), following Collins \textit{et al.} \cite{Collins2013}. 
We wish to emphasize that these results were obtained with air as the background fluid. 
However, this is not critical, as we will calibrate the model based on a particle-tracking experiment. 
The radius $R_d$ serves as a fitting parameter, allowing us to tune the numerical data and to compensate for model inaccuracies. 
Therefore, the value of $R_d$ represents the droplet size distribution in an average manner. 
Also, we consider an immobilized interface between the droplets and the continuous phase, such that the drag force acting on the droplets is given by the Stokes drag. 
Furthermore, we assume that the droplets move with their terminal velocity relative to the background fluid, which is determined by equilibriating the Stokes drag and electric force as
\begin{equation}
\vec{u}_d = \frac{0.44 \, q_R \vec{E}}{6 \pi \mu_o R_d} =0.587  \frac{ \sqrt{\epsilon _0 \gamma R_d} }{ \mu_o}\vec{E}. 
\end{equation}
The droplet velocity $u_d$ is now a function of the local electric field, liquid properties, as well as the droplet radius $R_d$.

The advection-diffusion equation for charge reads 
\begin{equation}
\label{eq:adv_diff_charge}
\frac{\partial \rho_\text{el}}{\partial t} 
+ \nabla \cdot \left( 
\left(\vec{u}_d + \vec{u}_o \right) \rho_\text{el}
- D_d \nabla \rho_\text{el}
\right) = 0,
\end{equation}
where the charge is advected relative to the background fluid with the droplet velocity $\vec{u}_d$, in addition to being transported passively with the velocity of the background fluid $\vec{u}_o$. Also, the equation contains a diffusive flux with diffusion coefficient $D_d$. This term serves as an \textit{ad-hoc} approach to stabilize the numerics by preventing oscillations. In order to ensure that the problem is advection-dominated, the diffusive timescale $t_\text{dif}$ has to be much larger than the advective scale. As an approximation, we can calculate the velocity of a droplet with a radius of $\SI{5}{\micro m}$ and an electric field strength of $(3000/0.05)\,\si{V \per \meter}$, matching the initial field strength in the gap. The resulting velocity of $\SI{19.9}{ \milli \meter \per \second}$ results in a travel time from electrode to interface of $t_\text{adv} \approx \SI{0.25}{\second}$. Choosing $t_\text{dif} = 100\, t_\text{adv}$ leads to a diffusion constant of $D_d = \SI{1e-6}{\meter^2 \per \second}$. 
In Appendix~\ref{sec:Supp_grid_convergence}, the dependency of the numerical solution on $D_d$ is analyzed, demonstrating that it was chosen sufficiently small to not influence the numerically obtained interface deflection.
In order to introduce charge into the system, a boundary current is applied at the upper electrode as 
\begin{equation}
i_{in} = \frac{I_0}{\pi R_n^2}, 
\end{equation}
where $R_n$ denotes the needle radius. At the lower interface, an outflow condition is prescribed, allowing the outflow of charge as 
\begin{equation}
i_{out} = -\left( \vec{u}_d \cdot \vec{n} \right) \rho_\text{el}, 
\end{equation}
where $\vec{n}$ denotes the interface normal. At the other walls, zero flux conditions are prescribed. 

The electrostatic problem is governed by Poisson's equation 
\begin{equation}
\label{eq:Poisson_eq}
\nabla^2 \phi = -  \frac{\rho_\text{el}}{\epsilon_0 \epsilon_o},
\end{equation}
where $\rho_\text{el}$ represents the electric charge density inside the oil phase due to the electrospray. For the purpose of the simulations, we treat it as a continuous interpenetrating medium. The equation is supplemented by fixed potential boundary conditions at the needle as well as the oil-water interface as 
\begin{equation}
\label{eq:BC_potential}
\phi_n = \SI{3}{\kilo V} \quad \text{and} \quad \phi_i = \SI{0}{\kilo V}.
\end{equation}
The latter is a reasonable simplification, due to the high conductivity of the KCl solution compared to the oil phase, and allows us to neglect charge transport inside the lower phase. At the remaining boundaries, we prescribe a vanishing electric flux as $\vec{n}\cdot (\epsilon_0 \epsilon_r \vec{E}) = 0$. 

The fluid flow is governed by the continuity equation for incompressible flow  
\begin{equation}
\label{eq:continuity_eq}
\nabla \cdot \vec{u} = 0
\end{equation}
and the Navier-Stokes equations 
\begin{equation}
\label{eq:NavierStokes}
\rho \frac{\partial \vec{u}}{\partial t} + \left( \vec{u} \cdot \nabla \right) \vec{u} 
= \nabla \cdot \left[ -p \boldsymbol{I} + \mu \left( \nabla \vec{u} + \left( \nabla \vec{u})^T \right) \right) \right] 
+ \vec{f}_g + \vec{f}_\text{el},
\end{equation}
where $p$ denotes the pressure, $\vec{u}$ the velocity field, $\mu$ the dynamic viscosity, $\vec{f}_g$ the gravitational volume force term, and $\vec{f}_\text{el} = \rho_\text{el} \vec{E}$ the electric force term. 
In order to compute the interface deformation, we utilize an interface tracking scheme as implemented in \textsc{Comsol}, which prescribes at the interface
\begin{subequations}
\begin{align}
\vec{u}_o &= \vec{u}_w \\
\vec{n}_1\cdot  \left\llbracket -p \boldsymbol{I}  + \mu \left( \nabla \vec{u} + \left( \nabla \vec{u})^T \right) \right)  \right\rrbracket &= \gamma \left( \nabla_t \cdot \vec{n}_o\right) - \vec{n}_o\nabla_t \gamma, \label{subeq:inttension}
\end{align}
\end{subequations}
where the subscripts $o$ and $w$ indicate oil and water phase, $\gamma$ indicates the interfacial tension, and $\vec{n}$ denotes a unit normal vector on the interface. The second condition (eq. \ref{subeq:inttension}) includes the effects of interfacial tension. In order to allow deformation of the interface directly at the side-wall of the container, a Navier-Slip condition in combination with a no-penetration boundary condition is prescribed as 
\begin{subequations}
\begin{align}
{u}_z &=-  \beta \frac{\partial u_z}{\partial r}, \\
u_r &= 0,
\end{align}
\end{subequations}
where the slip length $\beta$ is set to a value of $\beta =\SI{20}{\milli m}$. Since we are only interested in the steady-state solution, the value can be chosen freely. At the contact line, the contact angle is fixed to $\SI{90}{\degree}$. At the top and bottom wall as well as at the needle, no-slip boundary conditions are imposed as  
\begin{equation}
\vec{u}=\vec{0}.
\end{equation}
Additionally, at the upper right edge, a reference pressure is prescribed in order to prevent numerical oscillations, defined as $p_\text{ref} = 3 \rho_o g h$, where $\rho_o$ denotes the mass density of the dielectric liquid, $g$ the gravitational constant and $h$ the distance between the needle and the undeformed interface. 

We compute the solution by using a time-dependent solver, until a steady state is reached. We utilize an implicit solver based on a backward differentiation scheme (\textsc{BDF}) with variable time-stepping, where we only restrict the maximum time step to $\SI{0.5}{\second}$, and a direct solver (\textsc{MUMPS}). The domain is represented by a mesh consisting of 22217 triangular cells with quadratic basis functions for the pressure and velocities, cubic basis functions for the electric potential and linear basis functions for the charge density. To ensure grid independence, we have performed a grid convergence study (see Appendix \ref{sec:Supp_grid_convergence} for more details). 

In summary, the numerical model computes the electric potential, fluid flow, interface deformation as well as the charge transport self-consistently. 
We wish to emphasize that the interface deflection is a result of the background flow of the dielectric liquid, dominating the Maxwell stress at the interface, which is neglected for the following computations at the liquid-liquid interface.
As a next step, we compute the resulting dimple form for different droplet radii $R_d$, and compare the computed velocity field to a calibration measurement.

\subsection{Calibration measurements}
We measure the velocity field of the dielectric liquid by utilizing particle tracking velocimetry (PTV). The pin electrode of the rectangular cell, as described in Fig. \ref{fig:Exp_setup}(a) is replaced by a needle, and the particle motion is recorded by a long-distance microscope with a $6 \times$magnification at a frame rate of $\SI{10000}{fps}$. Hollow glass spheres (diameter $5 - \SI{90}{\micro \meter}$, \textit{Cosheric LLC}) are sparsely added to the dielectric liquid, such that the liquid properties are not significantly influenced. We have verified that the sphere movement is induced by the background fluid flow. When the voltage is applied without a volume flow, no significant particle movement is present. This excludes dielectrophoretic or electrophoretic forces as the cause of the motion. The experiment is performed at a voltage of $\SI{3000}{V}$, a volume flow of $\SI{20}{\micro \liter \per \minute}$, a KCl concentration of $\SI{0.1}{\milli M}$, and an oil viscosity of $\SI{5}{\centi St}$. The high-speed video is recorded over a time span of $\SI{0.15}{\second}$.

\begin{figure}[tb]
\centering
\includegraphics[]{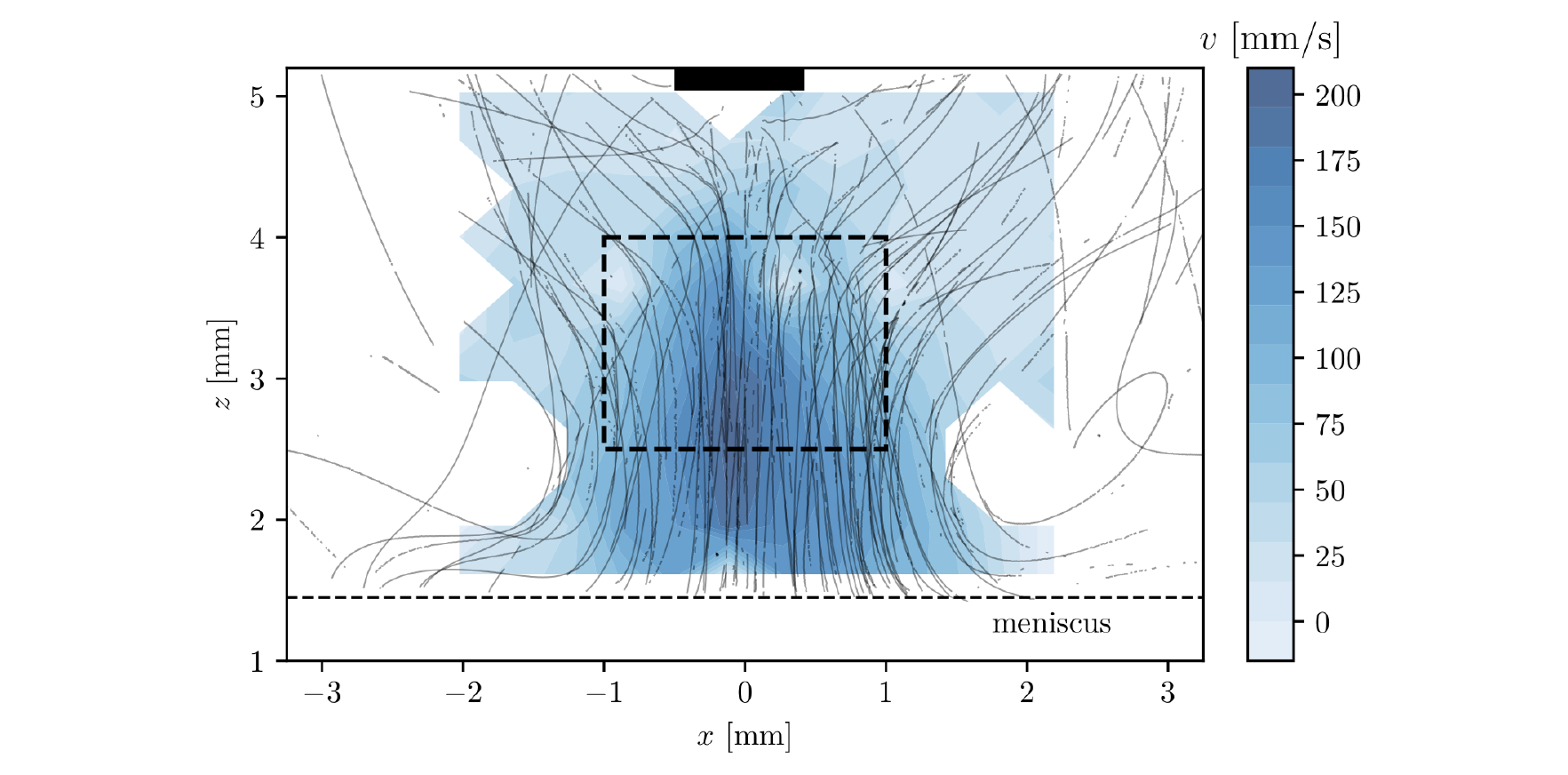}
\caption{\label{fig:numerics_calibration} 
Particle tracking velocimetry data obtained for $\SI{5}{cSt}$, $\SI{3000}{V} $, $\SI{20}{\micro \liter \per \minute}$ and $\SI{0.1}{mM}$ to calibrate the numerical model. Shown are particle streaklines and the magnitude of the $y$-component of the velocity field as colormap. The velocity is only displayed in regions with more than 30 velocity measurements. Close to the liquid-liquid interface, the velocities cannot be resolved due to the meniscus at the cell side wall. The needle is indicated at the top of the image. The dashed rectangle indicates the region of interest which we use to compare the experimental data with the numerical data. 
}
\end{figure}

Figure \ref{fig:numerics_calibration} shows the resulting $y$-component of the velocity field and the corresponding particle streaklines. We extract the velocities by utilizing the \textsc{Python} toolbox \textsc{trackpy} \cite{allan2014}, which computes the particle displacement between two images. Then, we define a grid of 20x15 cells ($54\times\SI{48}{px}$) and ensemble-average the $i$ velocities $v_i$, recorded within each cell during the measurement, as 
\begin{equation}
v = \sum_i^N \frac{1}{N} v_i. 
\end{equation}
For the final representation of the velocity field, we omit grid cells with less than 30 particle velocities $v_i$. As can be seen in Fig. \ref{fig:numerics_calibration}, the dielectric liquid assumes a significant vertical velocity of about $\SI{200}{\milli \meter \per \second}$, concentrated in a central plume below the needle, which is visible at the top of the image. The particle streaklines show that liquid is transported radially inwards towards the needle, where it accelerates. Then, close to the liquid-liquid interface, the streaklines are curved outwards. It is important to note that we are unable to resolve the velocities close to the interface, since the meniscus at the cell wall restricts our view. However, for the purpose of matching the velocities of the numerical computations, this region of observation suffices. In order to compare the experimental with the numerical velocity field, we average the vertical velocity field in a rectangular region $-\SI{1}{mm} \leq r \leq \SI{1}{mm}$ and $\SI{2.5}{mm} \leq z \leq \SI{4}{mm}$ (dashed rectangle in Fig. \ref{fig:numerics_calibration}), resulting in a value of $\SI{130}{\milli m \per \second}$.

\subsection{Numerical results}

\begin{figure}[tb]
\centering
\includegraphics[]{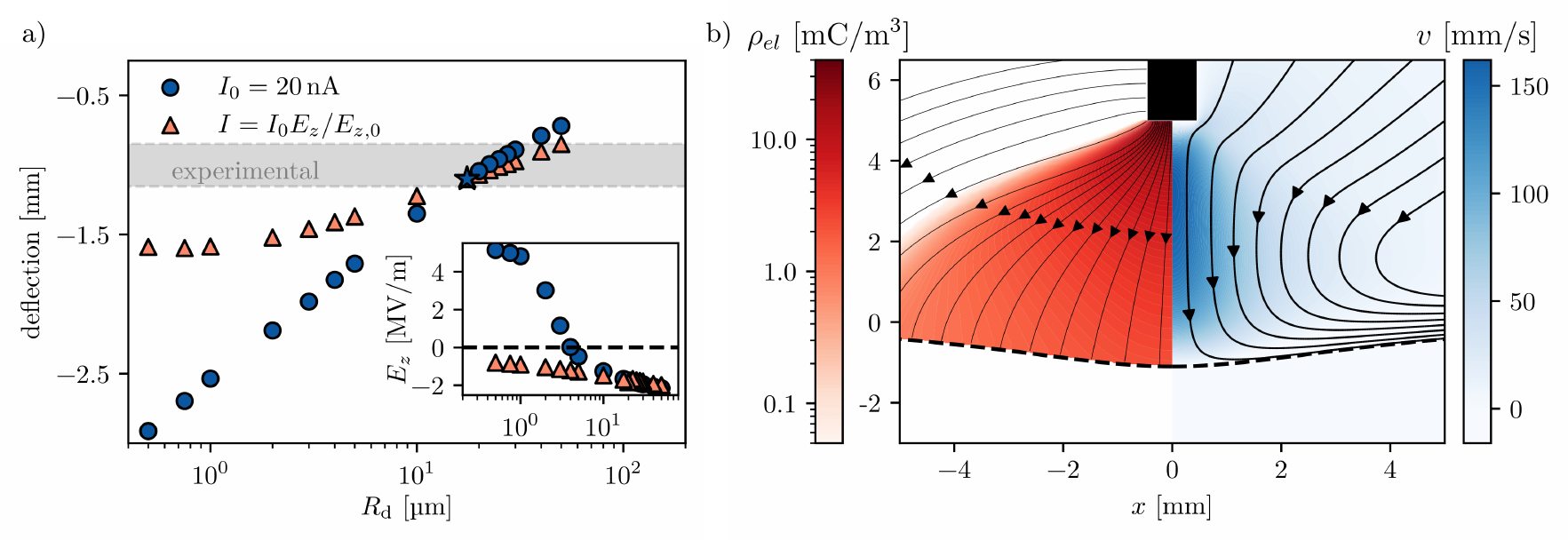}
\caption{\label{fig:numerics_results}
Results of the numerical model. 
a) The blue symbols depict the deflection of the liquid-liquid interface as a function of the radius $R_\text{d}$ for a fixed current $I_0$ and for similar conditions as in Fig. \ref{fig:numerics_calibration} (blue curve). The grey area depicts the experimentally observed deflection within the limits of two standard deviations around the mean value. At $R_d=\SI{17.5}{\micro m}$, the velocity field matches best between numerical calibration and calibration measurement, shown as a star symbol. By introducing a linear dependence of the current on the field strength (orange curve), the deflection reaches a limit for small droplets. Inset: The electric field strength at the center of the needle changes sign for small droplet sizes. 
b) Numerical results obtained for a droplet radius of $R_d=\SI{17.5}{\micro \meter}$ that shows the best agreement with experimentally observed velocities and deflections. The charge density (left) and magnitude of the $y$-velocity (right) are depicted, in addition to the electric field direction (left) and velocity streamlines (right).
}
\end{figure}

For a given current of $I_o = \SI{20}{\nano A}$, we calculate the resulting interface deflection as a function of the droplet radius $R_d$ ranging from $\SI{0.5}{\micro \meter}$ to $\SI{50}{\micro \meter}$. In Table \ref{tab:parameter_numerics} we summarize the parameters used in the numerical model, which correspond to the parameters used for the calibration measurement. As can be seen from Fig. \ref{fig:numerics_results}(a), the deflection increases exponentially with decreasing droplet size (note the logarithmic scale on the $x$-axis). For a droplet radius of $\SI{17.5}{\micro \meter}$, there is the closest match between the numerical velocity field and the experimental results. The comparison between numerical and experimental results was made based on the mean $y$-velocity within the region of interest depicted in Fig. \ref{fig:numerics_calibration}. The numerical computations yield an average velocity of $\SI{127}{\milli m \per \second}$, compared to $\SI{130}{\milli m \per \second}$ obtained by the calibration measurement. Figure \ref{fig:numerics_results}(b) shows the resulting space charge $\rho_\text{el}$ as well as the vertical velocity component $v$. The velocity reproduces the experimentally observed plume below the needle, and the fluid is replenished from above, as can be seen by the streaklines. The space charge distribution shows an accumulation in the central region, which provides a strong propulsion for the fluid. The numerical computations result in a deflection of the liquid-liquid interface of $\SI{1.1}{\milli \meter}$, while from the experiments a deflection of $\SI{1}{\milli \meter} \pm \SI{0.154}{\milli \meter}$ (mean value $\pm 2 \times \text{standard deviation}$) is obtained. Based on the fitted value of $R_d$, the numerically computed prediction for the interface deflection matches well with the corresponding experimental data. Therefore, we can be confident that the flow field induced by the charged droplets is responsible for the interface deflection.

In order to confirm that the Maxwell stress is dominated by hydrodynamic forces at the interface, we can use the simulations to compare it to the hydrodynamic pressure increase above the interface, as well as the pressure difference sustained at the interface by capillary forces.
For a droplet radius of $\SI{17.5}{\micro \meter}$, we observe a maximum pressure increase above the interface of \SI{18.64}{\pascal} compared to the case without an induced flow.
The interfacial tension sustains a maximum pressure difference at the interface of \SI{8.16}{\pascal}. The remaining pressure increase is balanced by the hydrodynamic pressure increase of the lower phase, as well as the hydrostatic pressure due to a shift in the interfacial position.
While the Maxwell stress is not included in the interfacial force balance, we can compute the resulting vertical electric stress at the interface as $\llbracket \sigma^M  \rrbracket \cdot \vec{e}_z$.
At the interface, the vertical Maxwell stress component shows a maximum value of \SI{1.42}{\pascal}, or \SI{17.4}{\percent} of the pressure difference sustained by interfacial tension.
From this \textit{a-posteriori} comparison, it follows that the interfacial deformation is governed mainly by the impinging background flow, with only a secondary effect of the Maxwell stress.

The computations with varying droplet radii for a given current yield noteworthy results for the vertical electric field $E_z$ at the center of the needle at $r=\SI{0}{\micro \meter}$, as can be seen in the inset of Fig. \ref{fig:numerics_results}(a). For droplets with a small radius, the electric field at this position changes its sign, which is clearly unphysical. When the electric field changes its sign, the electrospray current should change its sign as well. This artifact is a direct result of fixing the current at the needle as a boundary condition. If we replace this assumption by a linear dependency of the emitted current on the field strength at the needle, as was observed for example by Gundabala \textit{et al.} \cite{Gundabala2010}, the behavior changes drastically. When we use the electric field $E_{z,0}$ obtained in the case of $R_d = \SI{17.5}{\micro \meter}$ as the reference, we can prescribe the overall emitted current as $I = I_0 E_z({r=0,z=h})/E_{z,0}$. The resulting deflection shows a saturation for small droplet sizes. Now, the electric field no longer changes sign when the droplet radius is reduced. 
In the absence of a liquid-liquid interface, similar space charge regulation effects have been described  by Fernandez de la Mora \cite{delamora1992}, where the response of the liquid cone to the space charge was studied. Recently, a corresponding numerical study was presented \cite{Barrios-Collado2016}.
Also, a size segregation of the droplets was studied, where different droplet sizes separated along the radial coordinate with respect to the center line of the electrospray \cite{Ganan-Calvo1994}.
Furthermore, the self-regulation mechanism of the space charge is in line with observations made by \citeauthor{Larriba-Andaluz2010}, who sprayed an ionic liquid into a heptane bath \cite{Larriba-Andaluz2010}.
As we have hypothesized in section \ref{subsec:KCl_Concentration}, small droplet sizes lead to longer residence times, which in turn lead to a self-regulation of the electrospray by the space charge. Remarkably, we can reproduce the regulation mechanism using this comparatively simple numerical approach, and thus account for the experimentally observed limiting behavior.

\section{Conclusion and outlook}
\label{sec:conclusion}

Droplets have a significant influence on the dynamics of a liquid conductor-dielectric interface. They can be introduced during the electric-field induced interface disintegration of a present or past experiment. We use the results obtained for a steady-state system with controlled droplet injection into a dielectric liquid to rationalize the observations made for a pin-interface configuration (section \ref{sec:pin_interface_experiments}). Even a small flow rate at the needle can suffice to induce a significant background flow in the dielectric liquid \textit{via} \textit{viscous momentum exchange}. From the numerical computations it follows that especially droplets with a radius of the order of a few $\si{\micro \meter}$ have long residence times in the oil. Most likely, for the experiments depicted in Fig. \ref{fig:EHD_instabilities}(a-b), aqueous phase were deposited on the pin electrode in previous experiments, which then acted as a source of charged droplets. While the experiments presented here made use of liquids with moderate viscosities up to $\SI{10}{\centi St}$, this effect will become stronger with increasing viscosity. Also, as the density difference between dielectric and conducting liquid is usually small, sedimentation due to gravitation takes a long time, leading to droplet accumulation inside the dielectric liquid even over successive experiments. 

The electric field strength required to induce Taylor cones in EHD experiments is higher than what we have applied to create a steady-state dimple under prescribed volume flow. For example, interfacial breakdown in the experiments corresponding to Fig. \ref{fig:EHD_instabilities}(a) and (b) was observed around $\SI{4.5}{\kilo V}$ for an electrode gap of $\SI{4}{\milli \meter}$. At these higher electric fields, the force due to charged droplets in the oil phase becomes even more prominent than what we have observed here. The superposed effects of the induced flow field and the Maxwell stress at the liquid-liquid interface leads to intriguing phenomena. For example, while the background flow leads to an interface depression, the Maxwell stress induces Taylor cones from the rim of the dimple, as seen in Fig. \ref{fig:EHD_instabilities}(a) and (b). Additionally, surfactants shift the critical voltage for the occurrence of Taylor cones to smaller values. Simultaneously, they lead to a reduced interfacial tension, which increases the deformability of the liquid-liquid interface. As a result, phenomena like the formation of multiple cones might be triggered (Fig. \ref{fig:EHD_instabilities}(c)). Mapping the different regimes for varying surfactant concentrations is beyond the scope of this work and is left to future investigations. The same is true for the deeper exploration of the different dynamical regimes of the system that will depend on the interplay of different time scales such as the residence time of the droplets in the dielectric liquid and the characteristic capillary scale governing the interface deformation.  

Transient interactions between a Taylor cone and a space charge due to droplets can also lead to a self-limiting behavior. When droplets are ejected from a Taylor cone, they revert their charge upon contact with the pin electrode. As a result, they are repelled towards the liquid-liquid interface, and can suppress the Taylor cone, as for example shown in Fig. \ref{fig:EHD_instabilities}(d). Once the cone vanishes, no additional droplets are emitted into the system. After the droplets from the initial Taylor cone have merged with the lower phase, the Maxwell stress forms a Taylor cone again, leading to an overall oscillatory behavior. While in this case, the droplet influence is fairly obvious, it can be much more subtle, as shown in Fig. \ref{fig:EHD_instabilities}(a-b). While droplets are not immediately visible during the experiments, liquid residue at the electrode from previous experiments acts as a droplet source and leads to the surface dimple. 

In the parameter studies where the dimple in the interface was caused by the submerged electrospray, the voltage has a strong effect on the deflection, since it drives the droplet movement in the dielectric liquid. The volume flow, however, only shows a weak influence on both the electric current and the deflection. Increasing the dielectric liquid's viscosity leads to smaller currents and larger deflections, where for the highest viscosity of $\SI{10}{\centi St}$ a plateau was observed. A similar plateau can be observed when increasing the ion concentration in the droplets injected into the system. This is an interesting result, since a higher concentration corresponds to a higher electric conductivity, which has been associated with increasing currents in literature. However, by considering the decreasing size of the emitted droplets and the resulting space charge formed between the interface and the electrode, we identified an appropriate limiting mechanism. 

In order to fully account for the background flow, we utilize numerical simulations, which confirm that charged droplets induce a background flow and that this is the relevant mechanism of the interface deflection. In the charge transport equation, the droplet radius is an unknown parameter, which we determine by additional PTV measurements. By matching the velocity field between numerics and experiments, we recover the surface deflection as well. Also, the limiting effects of the space charge can be reproduced. These results corroborate that the background flow of the oil phase is indeed responsible for the observed interface dimple.

To summarize, we have investigated how droplets influence interfacial deformation modes and instabilities and have identified some key mechanisms governing the time evolution of the liquid-liquid interface. Because of the small size of the emitted droplets, the underlying phenomena can be easily overlooked, and thus lead to faulty interpretation of experimental data. We wish to especially highlight the importance of the flow inside the dielectric liquid in EHD experiments. Even for low viscosity fluids, the space charge due to droplets can lead to counter-intuitive results. We hope that our work will contribute to the overall understanding of EHD instabilities, and also to mitigation strategies where the effects of charged droplets are undesired. 

\begin{acknowledgments}
We thank M. Hartmann and F. Pl\"{u}ckebaum for the interfacial tension measurements. Financial support by the German Research Foundation (DFG) through Grant No. HA 2696/45-1 is gratefully acknowledged.
\end{acknowledgments}

\appendix

\section{Conductivity of KCl solutions}
\label{sec:Supp_KCl_cond}
The electric conductivity of aqueous KCl solutions is calculated from reference data \cite{shreiner2004}. We fit the relative conductivity $\Lambda$, which is defined as the conductivity per molar concentration, at $T=\SI{20}{\celsius}$ to the Kohlrausch law \cite{job2011}
\begin{equation}
\label{eq:Kohlrausch}
\Lambda = \Lambda^0 - A\sqrt{c},
\end{equation}
where $\Lambda^0$ denotes the relative conductivity at infinite dilution, and $A$ a fitting parameter. 
As a result, we obtain the conductivity data represented in Table \ref{tab:conductivity_KCL}. 

\begin{table}[b]
\caption{\label{tab:conductivity_KCL} Conductivity data for KCl solutions. }%
\begin{ruledtabular}
\begin{tabular}{ddd}
\multicolumn{1}{c}{KCl concentration $c_{KCl}$}&
\multicolumn{1}{c}{relative conductivity $\Lambda$}&
\multicolumn{1}{c}{Absolute conductivity $K$}\\
\multicolumn{1}{c}{ ($\SI{}{mol \per \meter^3}$)}&
\multicolumn{1}{c}{ ($\SI{}{\milli S \meter^2 \per  mol }$)}&
\multicolumn{1}{c}{($\SI{}{\milli S \per \meter}$ )}\\
\colrule
$0.1$ & $95.6$ & $9.6$ \\
$1$ & $93.4$ & $93.4$ \\
$10$ & $86.6$ & $866.5$ \\
$20$ & $82.5$ & $1650.5$ \\
\end{tabular}
\end{ruledtabular}
\end{table}

\begin{table}[t]
\caption{\label{tab:silicone_oil} Material parameters of the silicone oils as obtained from the data sheets and interfacial tension measurements.}%
\begin{ruledtabular}
\begin{tabular}{ddddd}
\multicolumn{1}{c}{kinematic viscosity $\nu$}&
\multicolumn{1}{c}{relative permittivity $\epsilon_r $}&
\multicolumn{1}{c}{density $\rho_o$} &
\multicolumn{1}{c}{surface tension $\sigma$}&
\multicolumn{1}{c}{interfacial tension to water $\gamma$}\\
\multicolumn{1}{c}{ ($\SI{}{\centi St}$)}&
\multicolumn{1}{c}{ ( - )}&
\multicolumn{1}{c}{($\SI{}{\kilo \gram  \per \liter}$ )}&
\multicolumn{1}{c}{($\SI{}{\milli N \per \meter}$ )}&
\multicolumn{1}{c}{($\SI{}{\milli N \per \meter}$ )}\\
\colrule
$0.65$ & $2.20$ & $0.76$ & $15.9$ & $39.6$\\
$1$ & $2.31$ & $0.92$ & $17.5$& $42.0$ \\
$5$ & $2.56$ & $0.92$ & $19.2$& $35.9$ \\
$10$ & $2.67$ & $0.94$ & $20.2$& $36.4$ \\
\end{tabular}
\end{ruledtabular}
\end{table}

\section{Material properties of silicone oils}
\label{sec:Supp_Liquids}

For the silicone oils, we obtain the relative dielectric permittivity, the density and surface tension from the data sheet \cite{Wacker2002}. The interfacial tension was measured using the Du No\"{u}y ring method (\textit{DCAT 25, dataphysics}). The oils were used as delivered. The material properties are listed in Table \ref{tab:silicone_oil}.  

\section{Numerical simulation parameters}
The parameters used in the numerical computations of section \ref{sec:numerical_modeling} are listed in Table \ref{tab:parameter_numerics}. 
\begin{table}[b]
\caption{\label{tab:parameter_numerics} Parameters used in the numerical computations. }%
\begin{ruledtabular}
\begin{tabular}{ccc}
parameter&description&value\\
\colrule
$h$ & interface-needle distance & $\SI{5}{\milli \meter}$ \\
$h_w$ & height water layer & $9h$ \\
$h_o$ & height oil layer & $3h$ \\
$R_c$ & radius container & $20h$ \\
$R_n$ & radius needle & $\SI{0.45}{\milli \meter}$ \\
$\rho_o$ & density oil & $\SI{910}{kg \per \meter ^3}$ \\
$\rho_w$ & density water & $\SI{1000}{kg \per \meter^3}$ \\
$\nu_o$ & kinematic viscosity oil& $\SI{5}{\centi St}$ \\
$\nu_w$ & kinematic viscosity water& $\SI{1}{\centi St}$ \\
$\epsilon_o$ & relative permittivity oil& $\SI{2.5}{}$ \\
$\gamma$ & oil-water interfacial tension & $\SI{36}{\milli N \per m}$ \\
$\beta$ & side wall slip length & $\SI{20}{\milli \meter}$ \\
$\Theta_c$ & side wall contact angle & $\SI{90}{\degree}$ \\
$\phi_n$ & needle potential & $\SI{3}{\kilo V}$ \\
$\phi_i$ & interface potential & $\SI{0}{\kilo V}$ \\
\end{tabular}
\end{ruledtabular}
\end{table}

\begin{figure}[t]
	        \centering{
	        \includegraphics[]{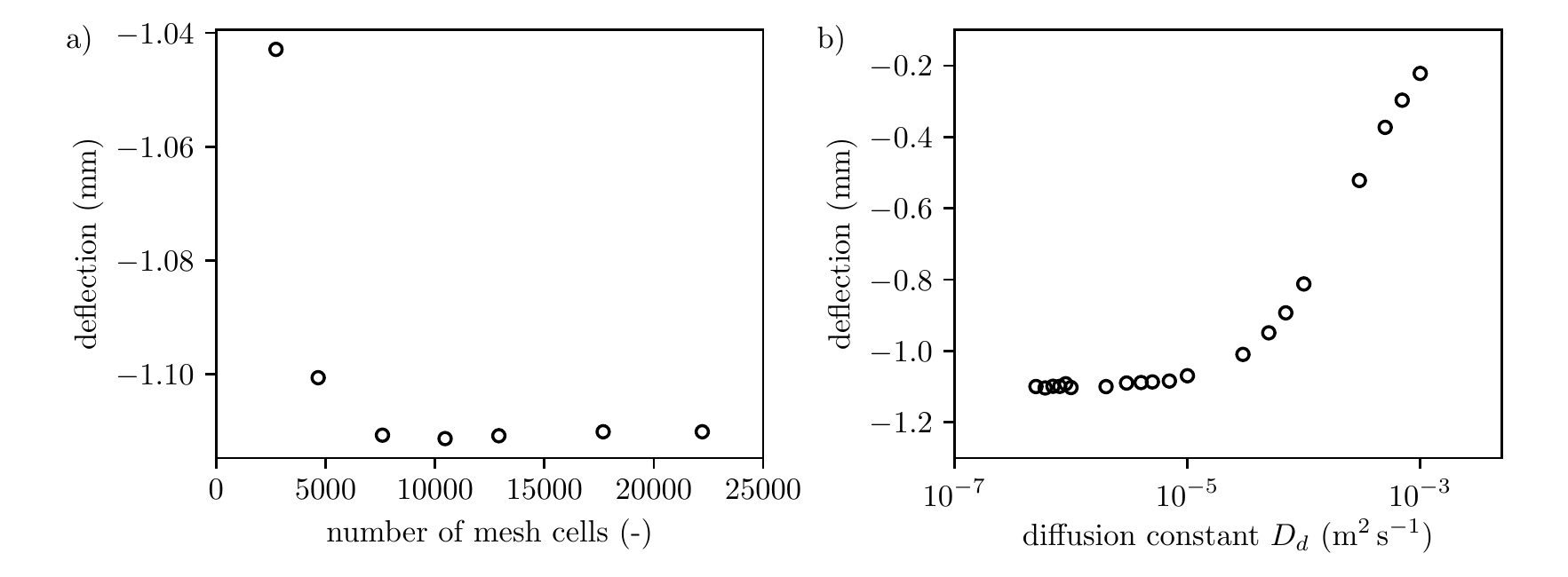}
	        \caption{Dependency of the solution on numerical parameters, expressed as the interface deflection as a function of these parameters.
	        a) Influence of the number of mesh cells. 
	        b) Influence of the diffusion constant $D_d$.       
	         }        
	        \label{fig:convergence}
	        }
\end{figure}

\section{Grid convergence study}
\label{sec:Supp_grid_convergence}

In order to ensure sufficient grid convergence in the numerical simulations, we analyze the influence of the cell size of the tetrahedral mesh on the interface deflection. 
For the case of $R_d=\SI{3}{\micro \meter}$, we vary the number of grid cells at the symmetry axis of the system, and limiting the maximum growth ratio between adjacent cells of $1.015$. As a result, the number of cells in the system changes. By this procedure, we add cells in the region of the highest importance without straining computational resources by adding cells in less important regions, e.g. in the conducting liquid far away from the symmetry axis. We considered seven different meshes. The number of mesh cells in the overall domain varies from $4661$ for the coarsest mesh to $22217$ for the finest mesh. For all different configurations, we computed the depth of the dimple, plotted versus the cell count in Fig.~\ref{fig:convergence}(a). It is apparent that the solution becomes virtually grid-independent for the highest spatial resolution. For the studies reported in the main text, we used the highest resolution to ensure that our solution is grid-independent.

In order to demonstrate that the solution is independent of the numerical \textit{ad-hoc} stabilization approach, the solution has to be advection-dominated, being independent of the diffusion constant $D_d$.
In Fig.~\ref{fig:convergence}(b), the interface deflection is shown as a function of the diffusion constant $D_d$, obtained for the droplet radius $R_d = \SI{17.5}{\micro\meter}$. For large diffusion constants, the deflection shows a strong dependence on $D_d$, as the charge transport is diffusion-dominated. Below a critical value, however, the solution becomes independent of the numerical value of $D_d$. 
For the final value of $D_d=\SI{1e-6}{\square\meter\per\second}$ used in the simulations, the deflection is independent of $D_d$, as the charge transport is advection-dominated.

\end{document}